\documentclass[aps,floatfix,superscriptaddress,notitlepage,nofootinbib]{revtex4-1}
\usepackage{amsmath,amssymb,amsfonts,dcolumn,bm,enumerate}
\usepackage{graphicx}
\usepackage{comment,natbib,appendix}
\usepackage{multirow,color}
\usepackage{afterpage}
\usepackage{xcolor}
\usepackage{amsthm}
\usepackage{natbib}
\usepackage{hyperref}

\newcommand{\cmp}
 {\affiliation{Saha Institute of Nuclear Physics, Kolkata 700064, India.}}
 \newcommand{\isi}
 {\affiliation{Economic Research Unit, Indian Statistical Institute, Kolkata 700108, India.}}
 \newcommand{\raghunathpur}
 {\affiliation{Department of Physics, Raghunathpur College, Raghunathpur, Purulia 723133, West Bengal, India.}}
 {
 \newcommand{\SRM}
 {\affiliation{Department of Physics, SRM University-AP, Andhra Pradesh - 522240, India.}}
 \newcommand{\SRMCSE}
 {\affiliation{Department of Computer Science and Engineering, SRM University-AP, Andhra Pradesh-522240, India.}}
 
\begin{document}
\title{Evolutionary Dynamics of Social Inequality and Coincidence of Gini and Kolkata indices under Unrestricted Competition}

 \author{Suchismita Banerjee}
 \email[Email: ]{suchib.1993@gmail.com}
 \isi

 \author{Soumyajyoti  Biswas}
 \SRM

 \author{Bikas K. Chakrabarti}
 \isi \cmp  
 
 \author{Sai Krishna Challagundla}
 \SRMCSE

 \author{Asim Ghosh}
 \raghunathpur
 
 \author{Suhaas Reddy Guntaka} 
 \SRMCSE

 \author{Hanesh Koganti}
 \SRMCSE

 \author{Anvesh Reddy Kondapalli}
 \SRMCSE

 \author{Raju Maiti}
 \isi

 \author{Manipushpak Mitra}
 \isi 
 
 \author{Dachepalli R. S. Ram}
 \SRM

\begin{abstract}
Social inequalities are ubiquitous and here we show that the values of the
Gini ($g$) and Kolkata ($k$) indices, two generic inequality indices, approach each
other (starting from $g = 0$ and $k = 0.5$ for
equality) as the competitions grow in various social institutions like markets,
universities, elections, etc. It is further showed that these two indices become equal and stabilize at a value (at $g = k \simeq
0.87$) under unrestricted competitions. We propose to view this coincidence of inequality indices as a generalized version of the
(more than a) century old 80-20 law of Pareto. Furthermore, the coincidence of the inequality indices noted here is very similar to the ones seen before for self-organized critical (SOC) systems. The observations here, therefore, stand as a quantitative support towards viewing interacting socio-economic systems in the framework of SOC, an idea conjectured for years.
\end{abstract}

\maketitle

\section{Introduction}
\label{sec1}
Human societies have evolved over the ages, and so did the organizations such as the economic (markets in particular), social (e.g., universities) and political
(parliaments or national assemblies) institutions. All
these systems are necessarily comprised of heterogeneous individuals competing for resources. Consequently, different
 measures of successes in such competitions have been developed. For example, inhomogeneous distributions of income or wealth (in markets),  majority or popularity (in elections), impact, award or citations (in universities, academic institutions, journals) are outcomes of `successes' in the respective contexts. One obvious and important feature in such systems has been the emergence of inequalities among the individual successes and their ubiquity \cite{stauffer}. Few are able to accumulate a large share of the resources, while the rest have a much smaller share. Interestingly, this unequal distribution is not always a consequence of finiteness of resources, but also happens in the cases where the resources are, for all practical purposes, limitless (say, citation distributions).  

\begin{figure}[!tbh]
\centering
\includegraphics[width=0.7\textwidth]{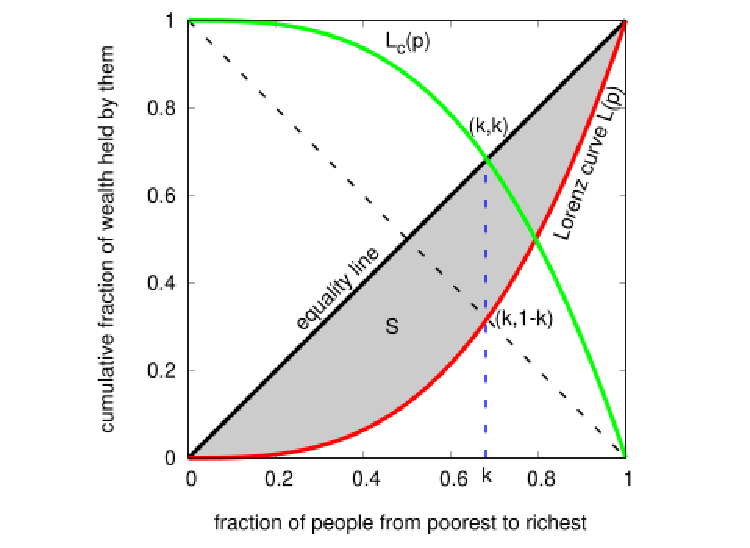}
\caption{The Lorenz curve (in red) $L(p)$ represents the fraction of overall income or wealth possessed by the bottom $p$ fraction of the people of a country (wealth can be replaced by vote counts or citations and people by the election-contestants or papers in case of elections or individual scientists respectively). The Gini coefficient is the ratio of the area that lies between the line of equality and Lorenz curve over the total area under the line of equality (Gini index $g = 2S$, $S$ represents area of shaded region). The complementary Lorenz function $L_c(p)\equiv 1-L(p)$ (see e.g., \cite{BCMM1}) is represented by the green line. The Kolkata index $k$  is given the ordinate value of the intersecting point of the Lorenz curve and the diagonal perpendicular to the equality line (implying $L_c(k) = k$, or $k$ corresponds to the fixed point of the complementary Lorenz function). The index value $k$ implies that $(1 - k)$ fraction of the richest  people possess $k$ fraction of the total wealth.}
\label{fig1}
\end{figure}

 Several quantitative measures of such social inequalities have been developed over the years. We will consider two such inequality measures, namely the old and widely used Gini index \cite{Gini1921} ($g$) and the recently introduced Kolkata index \cite{Ghosh2014} ($k$), both based on the Lorenz function or curve \cite{Lorenz1905} (see Fig. \ref{fig1}). The Gini index gives a measure of the average inequality among the population. Its range is $0 \leq g \leq 1$; where $g=0$ corresponds to perfect equality and $g=1$ corresponds to extreme inequality. On the other hand, the Kolkata index gives the fraction of wealth processed by $1-k$ fraction of the rich population (see e.g., \cite{BCMM1}) and, as such,  quantifies the 80-20 law ($k=0.8$) by Pareto \cite{Pareto1971Translation}. The range of the Kolkata index is $0.5 \leq k \leq 1$, where $k=0.5$  corresponds to perfect equality and $k=1$ corresponds to extreme inequality.

The formal definitions of these indices involve the so called Lorenz curve.
The Lorenz curve, in general, can be defined for any given distribution function $P(n)$, where the $n$th agent (a trader, a researcher and so on) has $P(n)$ fraction of the total resource (money or citation and so on) under consideration. 
The general procedure for obtaining the Lorenz function is the same for any set of agents possessing some (generally unequal) amount of resource. Let us take the specific example of papers written by an author and its citations.  
To get the Lorenz function for citation, the papers by the author can be arranged (on the $x$-axis) in an ascending order starting from the paper with least number of citation(s) to the paper with the highest number of citations (normalized by the total number of papers). Once such ordering of the papers is done, the associated cumulative fraction $L(p)$ of citations corresponding to the fraction $p$ of the least cited papers gives the Lorenz curve of that author (see Fig.~\ref{fig1}). One can then use this Lorenz curve to analyze the inequality in the citations of the author's published works and compare it with the corresponding inequality measures from the Lorenz curves of other authors. As mentioned before, similar analysis could be done for other quantities, for example a population could be arranged in the ascending order of the individual wealth and from the corresponding Lorenz curve the inequality among those individuals in the population could be quantified. The same applies for the vote shares of politicians, incomes from movies etc.

In this work, we find that across the social institutions (markets, universities, elections, etc), values of the Gini ($g$) and Kolkata ($k$) indices, estimated from the corresponding Lorenz curves as described above, approach each other as the competitions grow and they become equal (nearly $0.87$) under unrestricted competition. 
By unrestricted competition, we mean a competitive/interactive system devoid of any kind of external intervention intended specifically towards reducing the inequality across the participating agents. Examples may be, absence of social welfare measures and/or inequality reducing policies. Further, any form of subsidy (like unemployment dole) is absent for such instances of unrestricted competition. Ideal examples of unrestricted competitions include citation of papers, income from movies (Hollywood and Bollywood movies considered here), the Bitcoin market and electoral competition. Non-ideal examples include income/wealth among the population in a country (we have considered the data in the USA). The latter is important in view of the growing inequality and its possible limiting value in case when public welfare programs are consistently reduced. 

We analyze here the income data (both for income and income tax), in particular the Internal Revenue Service (IRS)  (USA) data \cite{IRS,Ludwig2021} for the period from 1983 to 2018, income from movie productions data both for Hollywood (USA \cite{Hollywood2011}) and Bollywood (India \cite{Bollywood2011}) films produced during the period 2011-2019, Google  Scholar citations data for papers written by scientists who won Fields Medal (mathematics), Boltzmann Medal (statistical physics), Dirac Medal (physics), and John von Neumann Medal (social science) in different years having individual Google Scholar pages with `verified email' address, and for vote share data of the candidates competing in parliament  elections in India for the last two election years (2014 and 2019 \cite{Lokesabha2014,Lokesabha2019}). All these data analysis  point towards the emerging coincidence of $g$ and $k$ indices to the value about $0.87$. 

It is important to note here that a set of interacting entities with no external fine tuning, can show emergent responses that are highly heterogeneous in nature. Such systems go under the common arch of self-organized criticality (SOC) \cite{soc,soc_bak}. In SOC systems, a collection of interacting entities can be infinitesimally slowly driven towards a critical point, where correlations diverge and consequently `bursty' responses are noted that are of all sizes (scale free). Such responses are governed by the characteristics (critical exponents) of the corresponding critical point, which behaves as an attractive fixed point for the system -- in stark contrast with tuned criticality where the critical point is necessarily repulsive. Due to the very fact that there is no fine tuning of the external parameters in SOC systems, the signature of an approaching critical point can only be seen in the corresponding scale free response. It has been shown very recently \cite{mbc} through numerical simulations for a group of generic SOC models (Bak-Tang-Weisenfeld, Manna sandpile models etc.) that the Gini ($g$) and Kolkata ($k$) indices of the inequality of their responses (in avalanche sizes in those cases) coincide to a value of about $0.87$ in the vicinity of the critical points. The approach of $g$ and $k$ towards each other near the critical point, therefore, serves as a remarkable indication of the approaching critical point.   

It has been argued for a long time \cite{soc25} that like in the case of some physical systems, interacting socio-economic systems also constitute  `complex systems' and can therefore show self-organized critical behavior \cite{zhukov}. Manifestation of such behavior could be seen in various different fields that include, but are not limited to, financial markets \cite{markets_soc}, Cryptocurrency markets \cite{bitcoin_soc}, citation indices \cite{cite_soc}, political behavior \cite{politics_soc} among many other dynamical systems. A similar connection between the multi-component socio-economic interacting systems could also be drawn from the standpoint of non-extensive entropy measures (Tsallis entropy \cite{tsallis1,tsallis2}), specifically through the formulation of the corresponding Lorenz curves by maximizing Tsallis entropy under the constraints of Gini index value \cite{lorenz_tsallis} for income distributions, through the identification of non-extensive triplets for crypocurrency markets \cite{bitcoin_tsallis}.

In general, for real-world data, such connections (to SOC and nonextensive statistics) are usually drawn from the scale free responses (e.g., distribution of wealth), specifically the fractal nature of the behavior of these systems at the largest scale (e.g., the wealthiest people), rather than from probing the dynamics of the system or the characteristics of the bulk of the individual agents. We show here that the behavior of the appropriately constructed Lorenz curve in general, and the inequality indices in particular, which can be argued to be measures of `average' responses rather than the extreme ones, show remarkable similarities with those examined \cite{mbc,front} in the case of physical models of SOC systems. The observations on a myriad of socio-ecoomic data presented here, therefore, provide a strong quantitative support of the parallels conjectured for a long time between SOC and socio-economic systems.  

We then also discuss some general analytical and structural features of the Lorenz function and the bounds for coincidence  values of the $g$ and $k$ indices. Generally speaking, we also observe here analytically the possible coincidence of the $g$ and $k$ values at around $0.87$. 

\section{Inequality analysis from data for income, income  tax, movie income, bitcoin fluctuation, citation and vote share: $g$ and $k$ indices}
\label{sec2}
As discussed above, various different socio-economic systems show emergent properties when their dynamics are not externally fine-tuned, resulting in an environment of unrestricted competition, in the present context. One example of such emegent properties is the prevalence of inequality. We argue that some universal features of such inequality is a consequence of the SOC nature of the underlying dynamics. It has extensively been shown before that many of these systems indeed show SOC dynamics (see e.g., Refs. \cite{zhukov,markets_soc,bitcoin_soc,cite_soc,politics_soc}). It has also been shown that SOC models exhibit near universal behavior in terms of the inequality of their responses \cite{mbc,front}. In this section, we analyze real data for some socio-economic systems and demonstrate that the near universal features of inequality that are seen in SOC models, are indeed present here.  

Specifically, we discuss the features of the inequality indices Gini  ($g$) and  Kolkata ($k$), using data for income, income tax  (IRS, US for the period 1983-2018 \cite{IRS,Ludwig2021}), Box office income from production of films in Hollywood (US, \cite{Hollywood2011}) and Bollywood (India, \cite{Bollywood2011}) for the period 2011 to 2019, the fluctuation of the Bitcoin values (v(t) measured in USD) taken from daily prices (during the period 2010-2021 \cite{Bitcoin2021}), citations of papers by a few international prize (Fields Medal, Dirac Medal, Boltzmann Medal and John von Neumann Award) winners who have individual pages ('verified email' address)  in Google Scholar and vote share in Indian parliament elections in 2014 and 2019.

As mentioned in the Introduction, the Gini and Kolkata indices are measured from the Lorenz curve. The Lorenz curve can be constructed from any finite series of data in the following way: First the series is to be arranged in the ascending order. After the arrangement, $L(p)$ denotes the ratio of the sum of the first $p$ fraction of terms in the series to the sum of the total series (see Fig. \ref{fig1}). Evidently, in a particular context, say when the series is the wealth of individuals, $L(p)$ denotes the fraction of the total wealth possessed by the poorest $p$ fraction of the population. Clearly, $L(0)=0$, $L(1)=1$. Also, when each term of the series is equal to one another, $L(p)=p$ -- a diagonal, often called the equality line. The other extreme is when only one term is finite and all others are zero, in which case $L(p)=0$ for $p\ne 1$, and $L(p=1)=1$. In all cases studied here, $L(p)$ lies in between these two extremes, a monotonically increasing function and is always below the equality line. The area opened up between the equality line and the Lorenz curve, therefore, is a measure of inequality. Indeed, the area between the equality line and the Lorenz curve, normalized by the area in the case of extreme inequality (the triangle joining (0,0), (1,0) and (1,1)), is the Gini index $g$. Formally, $g=1-2\int\limits_0^1L(p)dp$. The Kolkata index $k$ is also defined through the Lorenz function through the equation: $L(k)=1-k$. Clearly, its value says that $1-k$ fraction of the terms in the series sums up to $k$ fraction of the total sum of the series. In a particular example of the wealth distribution in a society, if the Kolkata index is $k$, then $1-k$ fraction of individuals possess $k$ fraction of the total wealth in that society. 

In what follows, we estimate these two indices from the real data of various different systems and look for their near universal characteristics (all our data collections and analysis were completed before the end of 2021).

\subsection{Inequality analysis of data for  income, income tax  and income from movies}
\label{subsecA}
\begin{figure}[!tbh]
\centering
\includegraphics[width=0.95\textwidth]{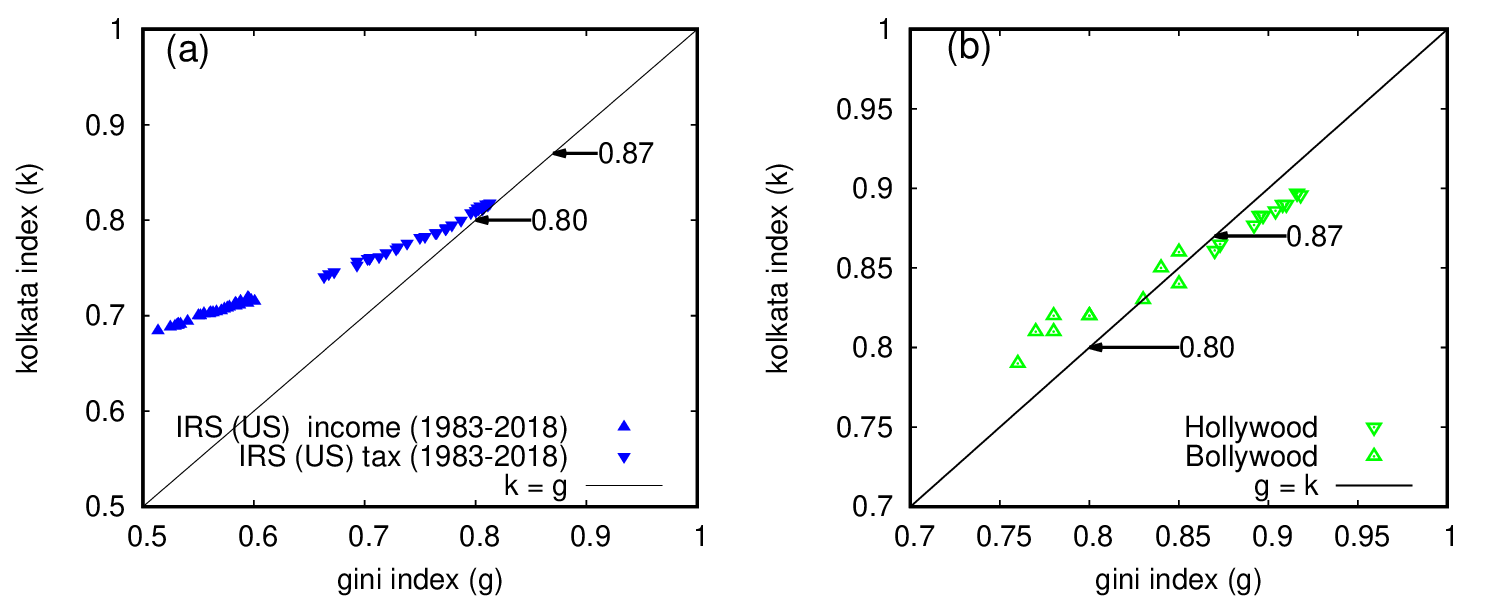}
\caption{(a) Plot of Kolkata index ($k$) against Gini index ($g$) for income  and income tax  extracted from IRS (USA) data \cite{IRS,Ludwig2021} for the period 1983 to 2018, from the corresponding Lorenz functions $L(p)$ for each of these 36 years. (b) Plot of Kolkata index ($k$) against Gini index ($g$) for (box office) income extracted from Hollywood (USA \cite{Hollywood2011}) and  Bollywood (India \cite{Bollywood2011}) for 9 years period 2011-2019.}
\label{fig2}
\end{figure}

We calculate the Lorenz function $L(p)$, representing the cumulative income by the poorest $p$ faction of people from the IRS (US) data for the income and as well as the income tax \cite{IRS,Ludwig2021} for 36 years (1983-2018). We then extract (e.g., Fig. \ref{fig1} and Ref.  \cite{BCMM1})  $g$ and $k$ indices values for each year (see Fig. \ref{fig2}a). We do the same for yearly (Box office) income from production of the films in Hollywood (USA \cite{Hollywood2011}) and  Bollywood (India \cite{Bollywood2011}) for 9 years period 2011-2019 (see Fig. \ref{fig2}b). 


\subsection{Inequality analysis of data for Bitcoins value fluctuations}
\label{subsecB}
\begin{figure}[!h]
\centering
\includegraphics[width=0.7\textwidth]{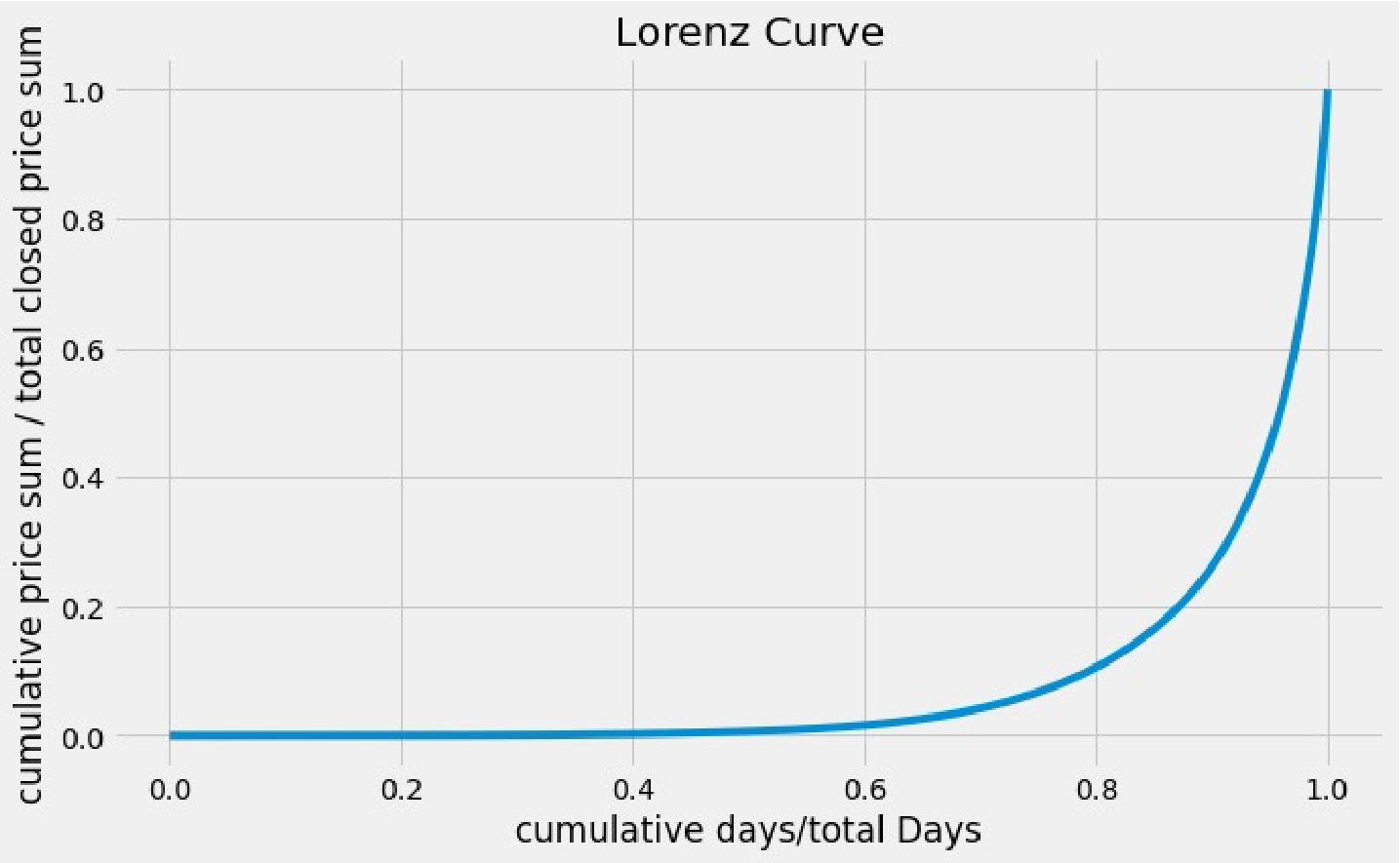}
 \caption{Lorenz curve for the difference of closing price of Bitcoins for successive days.}
\label{fig3a}
\end{figure}

\begin{figure}[!h]
\centering
\includegraphics[width=0.95\textwidth]{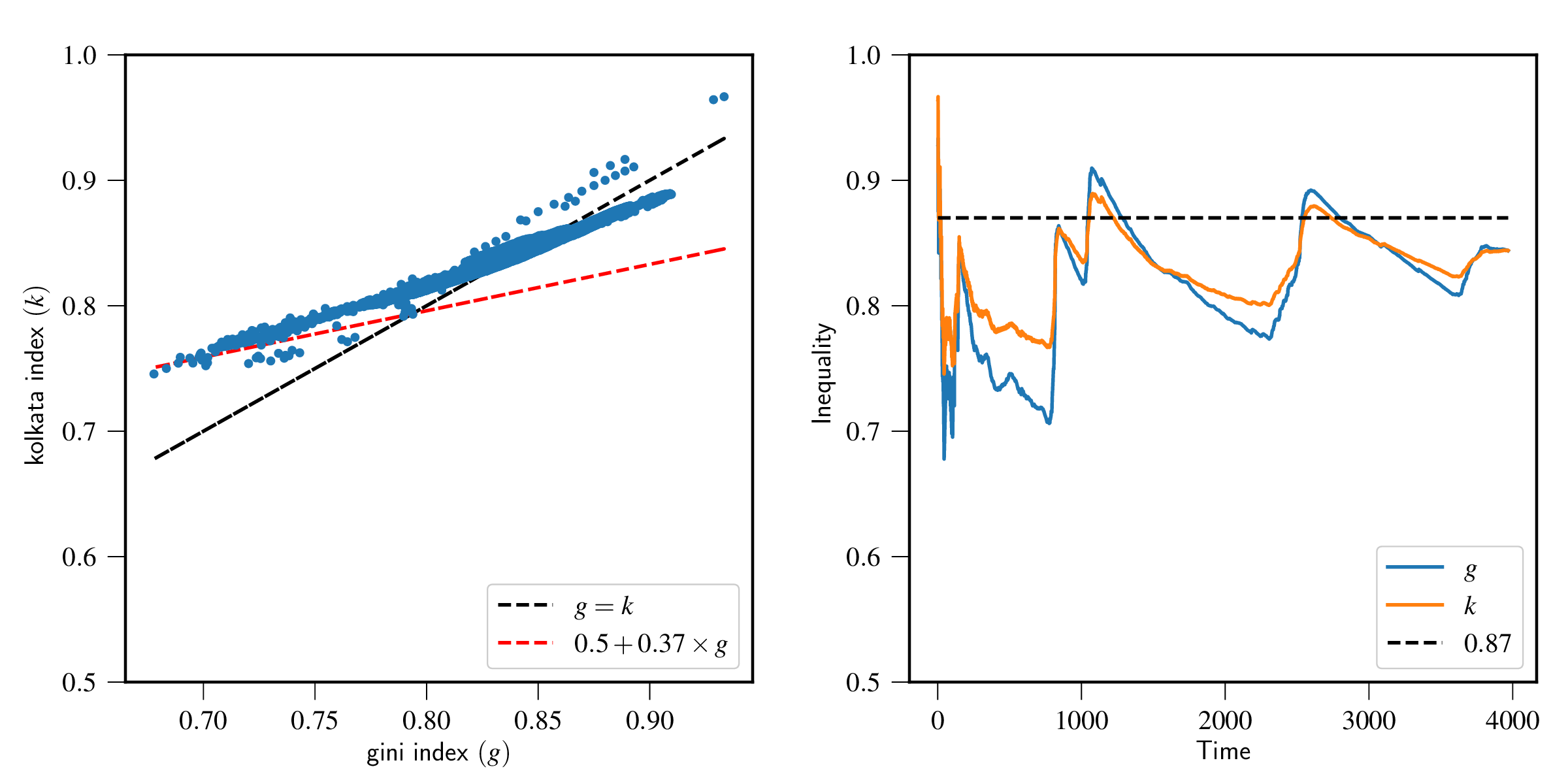}
 \caption{Left: Plot of Kolkata index ($k$) against Gini index ($g$) for the statistics of the bitcoin daily prices; Right: Temporal fluctuation of the $g$ and $k$ indices are shown. A reference value of $\simeq 0.87$ is indicated here for comparison.}
\label{fig3}
\end{figure}

Bitcoin is the first and the largest crypto-currency. Without any central bank to control, Bitcoin runs in a system of decentralized ledger. Introduced in 2008 (first used in 2009), the total market value of Bitcoin stands over \$1.03 trillion today (November, 2021), which is equivalent to about 2.9\% of the total narrow money supply of the entire world. The decentralized nature of Bitcoin has rendered to vulnerable to strong market volatility. On the other hand, however, this is an ideal example of unrestricted competition among the available currency markets. To look at the fluctuation of the Bitcoin values (v(t) measured in USD), we have taken daily prices during the period January 1, 2010 to November 24, 2021 \cite{Bitcoin2021}. After the bitcoin data set for daily closing price is collected, we calculated the absolute value of the price fluctuations in consecutive days and the fractional closing price changes are then collected. Leaving a few days (upto about $t_0$ of the O(10)), we get the Lorenz curve (see Fig.~ \ref{fig3a}) for the closing price data upto a date $(t>t_0)$ and proceed to estimate the Gini and k-indices as discussed in Fig.~ \ref{fig3}.

As can be seen from Fig. \ref{fig3}, the values of $g$ and $k$ repeatedly approach each other near 0.87. From an SOC standpoint, the critical point is an attractor and near that point $g_c=k_c\approx 0.87$ \cite{mbc}, which is what is reflected in the Bitcoin data here. 

\subsection{Inequality analysis of citations data for prize winners and vote data for election contestants}
\label{subsecC}
\begin{figure}[!tbh]
\centering
\includegraphics[width=0.7\textwidth]{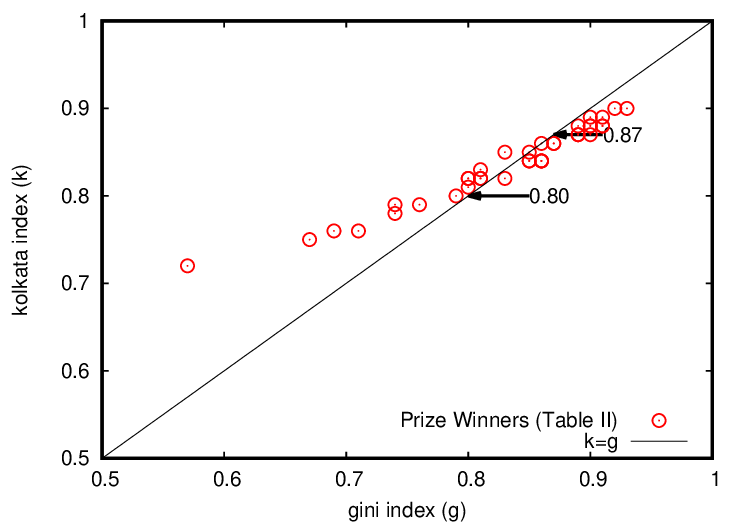}
\caption{Plot of Kolkata index ($k$) against Gini index ($g$) for the citation inequalities of the papers published by the individual prize winner (see table \ref{table2}),  extracted  from the corresponding Lorenz function $L(p)$ for each of the scientist.}
\label{fig5}
\end{figure}

We derive the Lorenz function $L(p)$ (representing the cumulative fraction of citations earned by $p$ fraction least cited papers) of citations by a few international prize winners (Fields Medal, Dirac Medal, Boltzmann Medal and John von Neumann Award) who have their own Google Scholar pages with `verified email' address (see table \ref{table2}). The result of the inequality analysis are shown in Fig.~ \ref{fig5}.
This is another ideal example of dynamics without external fine tuning or interventions. In this case also, $g$ and $k$ approach each other around 0.87. 

In the case of the vote shares of candidates in the last two general elections (2014 and 2019) of the Indian parliament, also show high inequality (Table \ref{table3}) and $g$, $k$ values are 0.83, 0.86 in 2014 and 0.85, 0.88 in 2019. Again, the proximity of these values to about 0.87 in the case of an unrestricted competition is seen.  

\begin{table}[!tbh]
 \centering
\caption{Citation data analysis for some of the selected Prize
winners in Physics (Dirac Medal \& Boltzmann medal) , Mathematics
(Fields Medal) and Social Sciences (John von Neumann Award), who have their respective
(`Verified email' and updated after 2018) Google Scholar pages
(data taken  in the 1st week of April 2021). Here, for each individual scientist, $N_p$
denotes total number of papers, $N_c$ denotes total
citations, $h$ denotes Hirsch index, $g$ denotes Gini
index and $k$ denotes Kolkata index.}
\label{table2}
\small
\resizebox{\columnwidth}{!}{%
\begin{tabular}{|c|c|c|c|c|c|c|}
\hline
Award & Name of recipients&\multicolumn{5}{c|}{Google Scholar citation data}\\
 \cline{3-7}
 &  & $N_p$ & $N_c$  & \multicolumn{3}{c|}{index values}\\
 \cline{5-7}
 & &  & & $h$ & $g$ &$k$\\
 \hline
 \multirow{10}{*}{\shortstack[lb]{ FIELDS \\ Medal \\ (Math.) }}  &Terence Tao&604&80354&101&0.88&0.86\\
\cline{2-7}
&Edward Witten&402&314377&212&0.74&0.79\\
\cline{2-7}
&Alessio Figalli&228&5338&41&0.67&0.75\\
\cline{2-7}
&Vladimir Voevodsky&189&8554&38&0.83&0.85\\
\cline{2-7}
&Martin Hairer&181&7585&48&0.74&0.78\\
\cline{2-7}
&Andrei Okounkov&134&10686&54&0.69&0.76\\
\cline{2-7}
&Stanislav Smirnov&79&4144&28&0.76&0.79\\
\cline{2-7}
&Richard E. Borcherds&61&5096&26&0.81&0.83\\
\cline{2-7}
&Ngo Bao Chau&44&1214&17&0.71&0.76\\
\cline{2-7}
&Maryam Mirzakhani&25&1769&18&0.57&0.74\\
 \hline 
 \multirow{9}{*}{\shortstack[lb]{ASICTP \\ DIRAC \\  Medal \\ (Phys.)}} & Rashid Sunyaev & 1789&103493&121&0.91&0.88 \\
\cline{2-7}
& Peter Zoller&838&100956&151&0.81&0.82\\
\cline{2-7}
& Mikhail Shifman&784&52572&107&0.85&0.84\\
\cline{2-7}
& Subir Sachdev&725&58692&109&0.83&0.82\\
\cline{2-7}
& Xiao Gang Wen&432&46294&95&0.80&0.82\\
\cline{2-7}
& Alexei Starobinsky&328&47359&89&0.81&0.82\\
\cline{2-7}
& Pierre Ramond&318&23610&57&0.89&0.87\\
\cline{2-7}
& Charles H. Bennett&236&89798&67&0.90&0.88\\
\cline{2-7}
& V. Mukhanov&208&27777&60&0.85&0.84\\
\cline{2-7}
& M A Virasoro&150&12886&35&0.90&0.87\\
\hline 
 
 \end{tabular}
 \begin{tabular}{|c|c|c|c|c|c|c|}
 \hline
 Award & Name of
recipients  & \multicolumn{5}{c|}{Google Scholar citation data}\\
 \cline{3-7}
  &  & $N_p$ & $N_c$  & \multicolumn{3}{c|}{index values}\\
 \cline{5-7}
 & &  & & $h$ & $g$ &$k$\\
 \hline 
 \multirow{5}{*}{\shortstack[lb]{BOLTZMANN \\  Award \\ (Stat. Phys.)}}&Elliott Lieb&755&76188&115&0.86&0.85\\
\cline{2-7}
&Daan Frenkel&736&66522&114&0.80&0.81\\
\cline{2-7}
&Harry Swinney&577&46523&92&0.86&0.84\\
\cline{2-7}
&Herbert Spohn&446&25188&78&0.79&0.80\\
\cline{2-7}
&Giovanni Gallavotti&446&15583&54&0.86&0.84\\
\hline
\multirow{10}{*}{\shortstack[lb]{JHON Von \\ NEUMANN \\ Award \\ (Social Sc.)}} &Daron Acemoglu&1175&172495&156&0.91&0.89\\
\cline{2-7}
&Olivier Blanchard&1150&113607&126&0.91&0.89\\
\cline{2-7}
&Dani Rodrik&1118&136897&145&0.90&0.89\\
\cline{2-7}
&Jon Elster&885&79869&109&0.89&0.87\\
\cline{2-7}
&Jean Tirole&717&201410&144&0.91&0.88\\
\cline{2-7}
&Timothy Besley&632&57178&90&0.89&0.88\\
\cline{2-7}
&Maurice Obstfeld&586&73483&94&0.90&0.88\\
\cline{2-7}
&Alvin E. Roth&566&54104&103&0.87&0.86\\
\cline{2-7}
&Avinash Dixit&557&82536&86&0.93&0.90\\
\cline{2-7}
&Philippe Aghion&490&119430&127&0.85&0.85\\
\cline{2-7}
&Matthew O. Jackson&397&39070&86&0.86&0.84\\
\cline{2-7}
&Emmanuel Saez&310&48136&75&0.86&0.86\\
\cline{2-7}
&Mariana Mazzucato&236&12123&44&0.87&0.86\\
\cline{2-7}
&Glenn Loury&226&13352&35&0.92&0.90\\
\cline{2-7}
&Susan Athey&203&18866&59&0.80&0.82\\
\hline 
 
\end{tabular}
}
\end{table}
 
 \begin{table}[!tbh]
 \centering
\caption{Gini ($g$) and Kolkata  ($k$) index values for the vote shares
in the last two elections of Indian Parliament in years 2014
and 2019 (number of contesting candidates in each year had been more
than 8000), obtained from the corresponding Lorenz function
$L(p)$, with data from refs. \cite{Lokesabha2014,Lokesabha2019}.}
\label{table3}
\begin{tabular}{|c|c|c|c|}
\hline
Year   &       Total voters  &      $g$   &        $k$  \\
\hline
2014    &      $5\times10^8$      &   0.83           &       0.86 \\
\hline
2019   &       $6\times10^8$         &  0.85          &       0.88 \\
\hline
\end{tabular}
\end{table}

\subsection{Near universal behavior of Gini  and Kolkata  indices}
\label{subsec-D}
\begin{figure}[!tbh]
\centering
\includegraphics[width=0.7\textwidth]{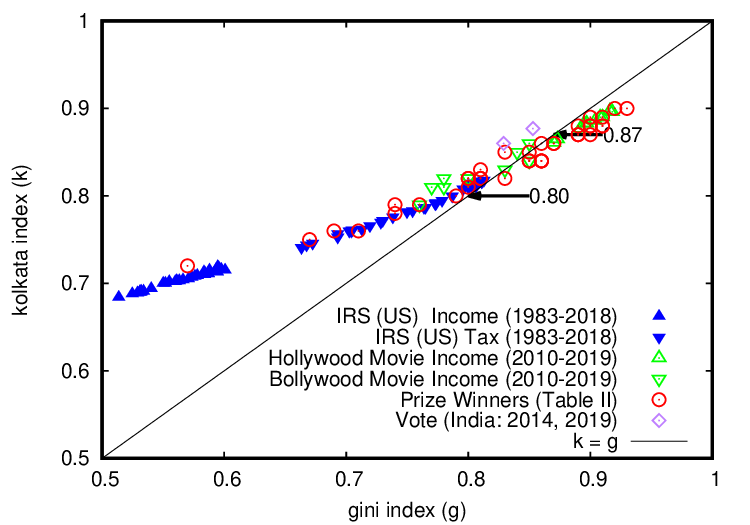}
\caption{A compiled plot of the Kolkata index ($k$) values against those of the corresponding Gini index  ($g$) for all the cases considered earlier in subsections ``Inequality analysis of data for  income, income tax  and income from movies'' (for household income and income tax data, Fig \ref{fig2}(a),  and movie income, Fig \ref{fig2}(b)) and ``Inequality analysis of citations data for prize winners and vote data for election contestants'' (for citation inequalities of the papers published by the individual prize winner, see table \ref{table2}, Fig. \ref{fig5} , and  vote share inequalities among the election contestants, see table \ref{table3}).  The over all fit suggests universal inequality behavior across the social institutions, and
the inequality convergence towards $k = g = 0.87\pm 0.02$.}
\label{fig6}
\end{figure}

We now try to compile all the results for the  Gini ($g$) and Kolkata ($k$) indices, obtained in earlier two subsections. We collect all the $g$ and $k$ estimate of subsection \ref{subsecA} from the IRS (US) data for household income and income tax during 1983-2018, and in subsection \ref{subsecC} from the Google Scholar citation
data for the papers published by 40 international prize winners  (Fields medalists, ASICTP Dirac medalists, Boltzmann medalists and von Neumann awardees, see table \ref{table2}) and also from the vote share data in the Indian Parliament elections in 2014 and 2019 (see Table \ref{table3}). These compiled results are shown in Fig. \ref{fig6}.  The overall fit of all these results clearly suggests universal growth of inequality across all the social institutions, markets (income, wealth), universities and academic institutions (citations), elections (vote shares among the election candidates), and the inequality measures convergence towards $k = g = 0.87\pm0.02$.

\begin{figure}[!tbh]
\centering
\includegraphics[width=0.7\textwidth]{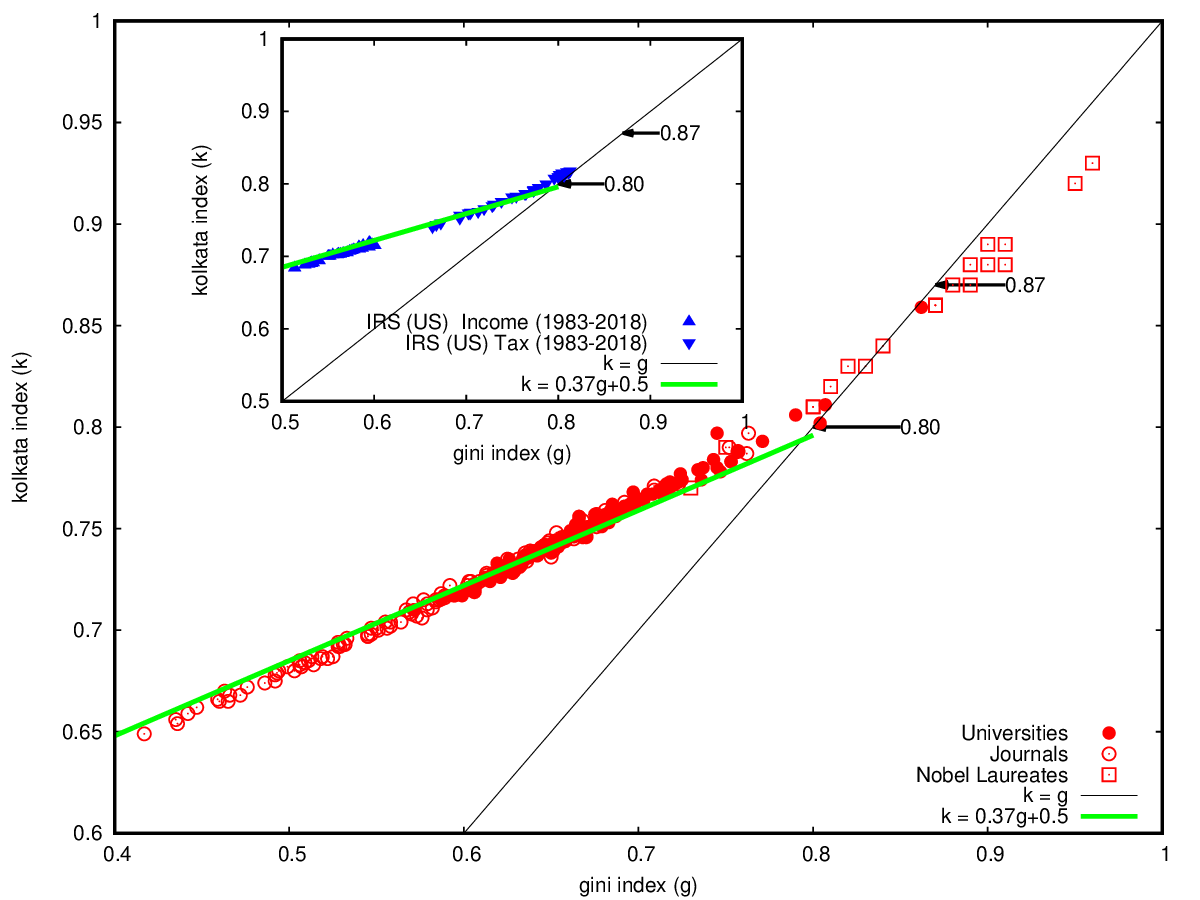}
\caption{ Comparison of $g$ and $k$ indices obtained here from IRS (US) data from 1983 to 2018 (from subsection ``Inequality analysis of data for  income, income tax  and income from movies''; see the inset) and (in the main fig) for the citations of papers published by  scientists from universities or institutes (results taken from Ref. \cite{Chatterjee2017}),  published in journals (results taken from Ref. \cite{Chatterjee2017}) and  20 Nobel Laureates in Physics, Chemistry, Medicine and Economics (results taken from Ref.  \cite{Ghosh2021}). Initial variation of $k$ against  $g$ for both income and income tax and for citations by universities, journals and individual scientists compared remarkably well and agree even quantitatively.}
\label{fig7}
\end{figure}

 In Fig. \ref{fig7}, we show how the growth pattern of the IRS (US) income inequalities ($k$ and $g$ values) compare with the growth of citation inequalities for the papers published by established universities, individual Nobel laureate scientists, and published in established journals (data taken from other publications, 
refs.\cite{Ghosh2021,Chatterjee2017,Chatterjee2016}). Again, all these results clearly indicate that the growth  of the values of the Kolkata ($k$) and Gini ($g$) inequality indices drift linearly towards a universal value of $k = g \simeq 0.87$  under unrestricted competitions, suggesting that about $87\%$ wealth, citations or votes are possessed, earned or won by $13\%$ people, papers or election candidates.

\begin{figure}[!tbh]
\centering
\includegraphics[width=0.7\textwidth]{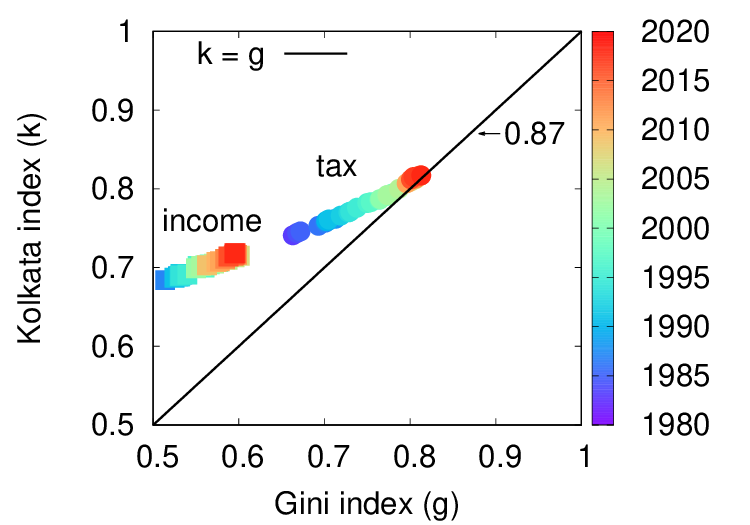}
\caption{Growths of $g$ and $k$ (USA economy, IRS data \cite{IRS,Ludwig2021}) shown with the variation in time (year). Clearly, the inequality measures grow with time, indicating a reduction in public welfare leading the system towards urestricted competition i.e., an SOC state. The value of $k$ in the tax data, which can be argued to represent the prevailing inequality status better, grows past the Pareto value (0.80) and predicted to reach 0.87 like in other the systems (e.g., movie income or citations) with complete absence of public welfare programs.}
\label{irs_gk}
\end{figure}
\subsection{Growth of inequality ($g$ and $k$) with decrease in welfare interventions from the IRS data}
So far we discussed the statistics of
the Gini ($g$) and Kolkata ($k$) indices
and about the observed terminal value of
both the indices, where $g$  becomes equal
to $k$ at around a value 0.87 (somewhat
higher than the value of $k$ = 0.80, as
suggested by the 80-20 law of Pareto).
In this subsection we discuss how the
same IRS income and tax data \cite{IRS, Ludwig2021} for
for 36 years (1983-2018) analyzed
in the subsection 2.1 (see also Fig. 2(a))
shows a consistent growth in the values of
the inequality indices $g$ and $k$ towards
a value 0.87 (see Fig. \ref{irs_gk}).

Here the year-wise growth of $g$ and $k$
obtained from analyzing the IRS income and
income tax  data, though goes past the
Pareto value of $k = 0.80$, does not  quite
reach  up  to  the value 0.87. It may be
noted that the income tax return data
(reflecting the income distribution among
the high income end) shows the inequality
index values to reach the maturity level
more prominently. As mentioned in the
Introduction, a possible reason could be
the social welfare interventions
(specifically  intended to reduce the
inequality  among the poorer sections)
are not directly affecting the competition
the richer section. The  other example,
the producers' income from movie productions
(discussed also in subsection 2.1)  show
also similar inequality level as there is
hardly any such  interventions.

\section{Some analytical obvservations regarding emerging coincidence of Gini and  Kolkata indices} 
\label{sec3}
The Gini coefficient for any distribution $F$ and Lorenz function $L(.)$ is given by $g=1-2\int_{0}^1L(p)dp$ and the Kolkata-index (or $k$-index) is given by that proportion $k\in [1/2,1)$ such that $k+L(k)=1$. Therefore, given any distribution function $F$, the coincidence of the Gini coefficient $g$ with the $k$-index $k$ takes place if and only if $2\int\limits_{0}^1L(p)dp=1-g=1-k=L(k)$. In the next three subsections we discuss more on the coincidence for different families of Lorenz functions.
\subsection{Non-trivial and symmetric Lorenz functions and the coincidence:} 
\begin{figure}[!tbh]
\centering
\includegraphics[width=0.95\textwidth]{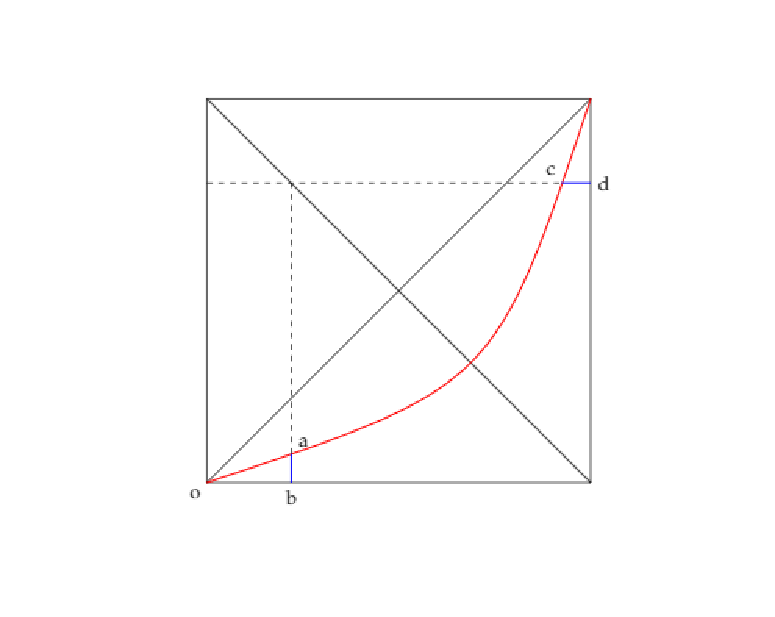}
\caption{Comparison of $g$ and $k$ indices obtained here from IRS (US) data from 1983 to 2018 (from subsection ``Inequality analysis of data for  income, income tax  and income from movies''; see the inset) and (in the main fig) for the citations of papers published by  scientists from universities or institutes (results taken from Ref. \cite{Chatterjee2017}),  published in journals (results taken from Ref. \cite{Chatterjee2017}) and  20 Nobel Laureates in Physics, Chemistry, Medicine and Economics (results taken from Ref.  \cite{Ghosh2021}). Initial variation of $k$ against  $g$ for both income and income tax and for citations by universities, journals and individual scientists compared remarkably well and agree even quantitatively.}
\label{symmetryLorenzF}
\end{figure}

A Lorenz function $L(p)$ is {\it non-trivial} if there exists $0<p<1$ such that $L(p)>0$. Non-triviality rules out the possibility that $g=1$ and hence ensures that $g\in [0,1)$. The Lorenz function $L(p)$ is \emph{symmetric} if for all $p\in (0,1)$ with $L(p)>0$,
\begin{equation}
\label{kpietra}
L(1-L(p))=1-p \ {\rm or} \ {\rm equivalently} \ L(p)+r(p)=1,
\end{equation}
where $r(p)=L^{-1}(1-p)$. The idea of symmetry is explained in Figure \ref{symmetryLorenzF}.

The simplest example of a non-trivial and symmetric Lorenz function is the Lorenz function associated egalitarian  distribution across agents, that is, with $L_1(p)=p$ for all $p\in [0,1]$. Since $L_1(p)>0$ for all $p\in (0,1)$, it is non-trivial. Moreover, for any $p\in (0,1)$, $L_1(1-L_1(p))=L_1(1-p)=1-p$. It was established in Banerjee, Chakrabarti, Mitra and Mutuswami \cite{BCMM1} that if a Lorenz function for some distribution $F$ is differentiable and symmetric, then the $k$-index coincides with the proportion associated with the Pietra index, that is, for any such distribution $F$, $k=F(\mu)$ where $F(\mu)$ is that cumulative proportion associated with the mean $\mu$ of the distribution $F$. For example, if we generalize the Lorenz function associated with the egalitarian distribution by consider any positive integer $m$ and any distribution $F_m$ such that the associating mean is $\mu_m$ and the associated Lorenz function $L_m(p)=1-\left(1-p^m\right)^{\frac{1}{m}}$, then one can easily verify that for any positive integer $m$ the Lorenz function $L_m(p)$ is non-trivial, symmetric and differentiable and $k_m=F_m(\mu_m)=(1/2)^{1/m}$. In this context, we have the following result.  
\begin{enumerate}
	\item[{\bf (A)}] \emph{If $F^*$ is any distribution function that generates a non-trivial and symmetric Lorenz function $L^*(p)$ such that the Gini coefficient $g^*$ coincides with the $k$-index $k^*$, then the coincidence value must necessarily belong to the interval $[3/4,1)$, that is, $k^*=g^*\in [3/4,1)$.}
\end{enumerate}
Result {\bf (A)} is proved in Appendix {\bf A}. A sufficiency result in this context is that for any rational  fraction $q\in \{3/4\}\cup [4/5,1)$, there exists a non-trivial and symmetric Lorenz function $L_q(p)$ such that $k_q=g_q=q$.  This result is elaborated in the next two paragraphs.

Consider the following Lorenz function:
\begin{equation}\label{coincidence34}
L_{\left(\frac{3}{4}\right)}(p)=
\left\{ \begin{array}{ll}
0 & \mbox{if $p\in \left[0,\frac{1}{2}\right]$,} \\
p-\frac{1}{2}  & \mbox{if $p\in \left(\frac{1}{2},1\right)$,} \\ 
1  & \mbox{if $p=1$.} 
\end{array} 
\right. 
\end{equation}
It is established in Appendix {\bf B} that the Lorenz function $L_{\left(\frac{3}{4}\right)}(p)$ (see Fig.~\ref{fig:lor}) given by (\ref{coincidence34}) is non-trivial and symmetric. Moreover, it is also established in Appendix {\bf B} that $L_{\left(\frac{3}{4}\right)}(p)$ given by (\ref{coincidence34}) is continuous, non-decreasing and convex in the open interval $(0,1)$ and that $k_{\left(\frac{3}{4}\right)}=g_{\left(\frac{3}{4}\right)}=3/4=0.75$. 

\begin{figure}[!tbh]
    \centering
    \includegraphics[width=0.7\textwidth]{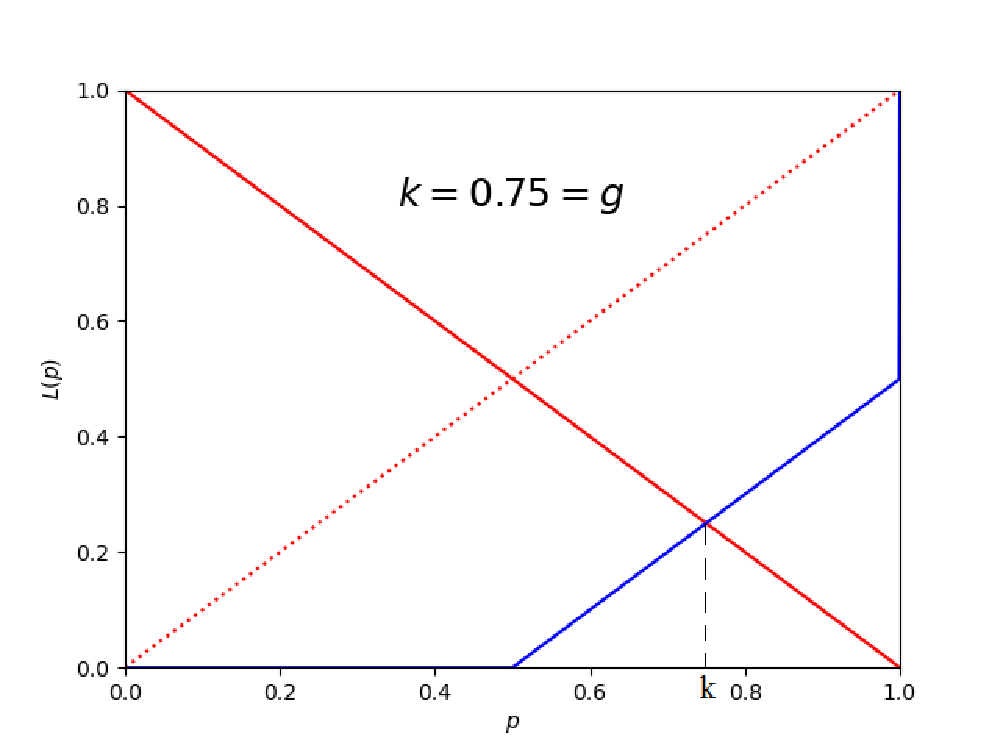}
    \caption{ The Lorenz curve $L(p)$ given by Eq.~(\ref{coincidence34}) is represented here by this blue line.
Red dotted line represents the equality line (i.e, $L(p) = p$) and the point $k = 0.75$ on the x-axis represents the $k$-index ($k$) value.Here Gini index ($g$) is the area between this Lorenz curve (blue solid line) and the equality line (red dotted line) normalised by the area under the equality line.For this Lorenz curve, here $g = 0.75 = k$. }
\label{fig:lor}
\end{figure}

\begin{figure}[!tbh]
\centering
\includegraphics[width=0.7\textwidth]{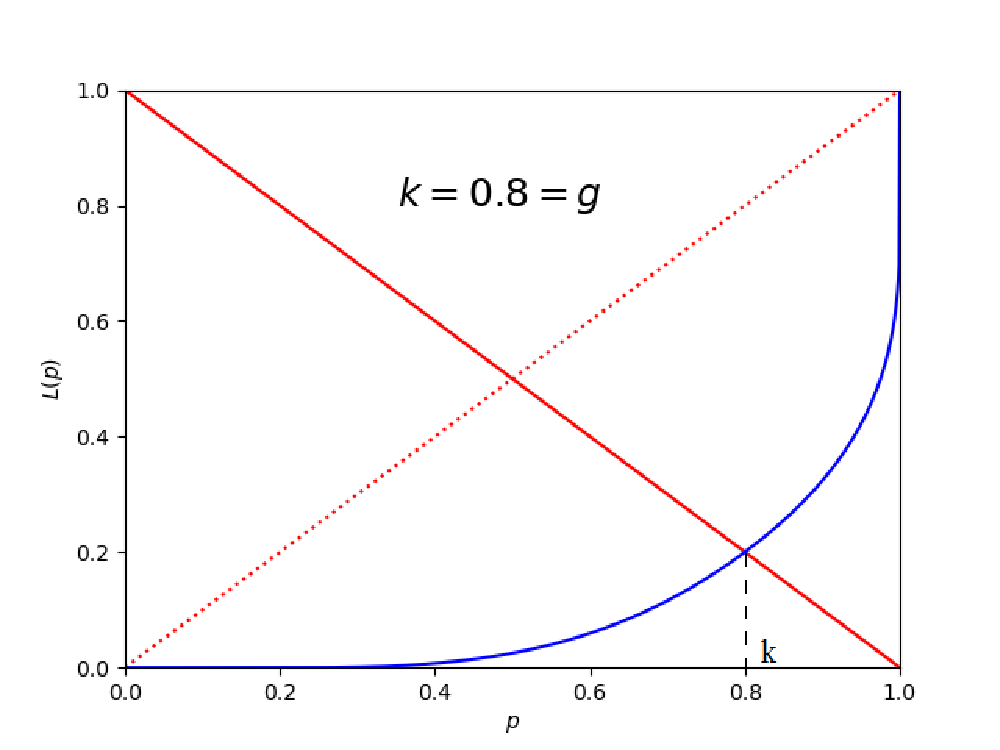}
\caption{The Lorenz curve $L(p)$ given by Eq.~(\ref{coincidencepoint8onwards}) for the special case $q = 4/5 = 0.8$ is represented here by the blue curve. Red dotted line represents the equality line (i.e, $L(p) = p$) and the point $k = 0.8$ on the x-axis represents the $k$-index ($k$) value. For this Lorenz curve, here $g = 0.8 = k$.}
\label{fig:lor_1}
\end{figure}

Next, we provide a family of non-trivial and symmetric Lorenz functions for which the Gini coefficient coincides with the $k$-index. Consider the following family of Lorenz functions(in Fig.~\ref{fig:lor_1}) defined for any rational fraction $q\in \left[\frac{4}{5},1\right)$:

\begin{equation}\label{coincidencepoint8onwards}
L_{(q)}(p)=
\left\{ \begin{array}{ll}
0 & \mbox{if $p\in \left[0,1-q\right]$,} \\
(1-q)\left(\frac{p+q-1}{2q-1}\right)^3  & \mbox{if $p\in \left[1-q,q\right]$,} \\ 
q-(2q-1)\left(\frac{1-p}{1-q}\right)^{\frac{1}{3}}  & \mbox{if $p\in \left[q,1\right)$,} \\
1  & \mbox{if $p=1$.} 
\end{array} 
\right. 
\end{equation}

It is established in Appendix {\bf C} that for every rational fraction $q\in [4/5,1)$, the resulting Lorenz function $L_{(q)}(p)$ given by (\ref{coincidencepoint8onwards}) is non-trivial and symmetric and that $k_{(q)}=g_{(q)}=q$. It is interesting to note that by setting $q=4/5=0.8$ in (\ref{coincidencepoint8onwards}) we get Pareto's 80/20 rule as a special case of coincidence between the Gini coefficient and the $k$-index. 

The Lorenz functions given in (\ref{coincidence34}) and (\ref{coincidencepoint8onwards}) are not analytic functions and we believe that it is this non-analyticity that is allowing for the coincidence between $k$-index and the Gini coefficient to have values in the interval $[3/4,1)$. We believe that with analytic functions, this coincidence value is bounded above by $8/9\simeq 0.889$. In the next three subsections we deal with Lorenz functions that are analytic and look for coincidence possibilities. 
 
\subsection{An exponential family of Lorenz functions}
Consider the Lorenz functions of the form $L_{\beta}(p)=\frac{(e^{\beta p}-1)}{(e^{\beta}-1)}$ for all $p\in [0,1]$ with $\beta>0$. Then in Appendix-({\bf D}) we first show that $g_{\beta}=\left\{\left(\beta-2\right)/\beta\right\}+\left\{2/\left(e^\beta-1\right)\right\}$. In Appendix-({\bf D}) we also show that the only solution to the coincidence between the Gini coefficient and $k$-index occurs at $\beta^*=14.778$ and as a result we get 
\begin{equation}
g_{\beta^*}=k_{\beta^*}=0.865\in \left(\frac{6}{7},\frac{13}{15}\right).\hspace{3in} 
\end{equation} 

\subsection{Two important families of Lorenz functions and a resulting hybrid family}
\subsubsection{The first family} 
\begin{figure}[!tbh]
\centering
\includegraphics[width=0.7\textwidth]{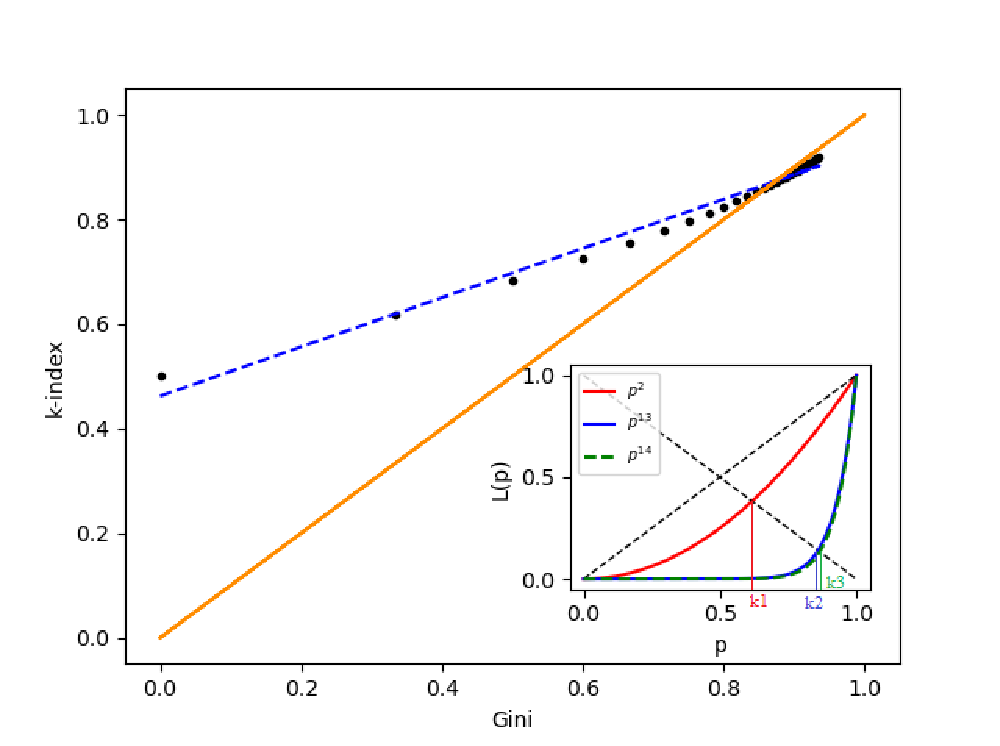}
\caption{Gini ($g$) vs. $k$-index ($k$) graph. Here Orange line represents the $g = k$ line. Black dots corresponds to ($g,k$) values for the Lorenz function = $p^n$ with $n\in\{1,20\}$. Blue dashed line is the best-fit line. In the inset red curve depicts the Lorenz curve for $L(p)=p^2$ and $k$-index value, k1 = 0.618, blue curve is the Lorenz curve for $L(p)=p^{13}$ and $k$-index value, k2 = 0.860 and similarly green dashed curve represents the Lorenz curve for $L(p)=p^{14}$ and $k$-index value, k3 = 0.866.}
\label{fig:lor_2}
\end{figure}

Consider another family of Lorenz functions of the form $L_N(p)=p^{N}$ (see Fig.~\ref{fig:lor_2}) for all $p\in [0,1]$ where $N\geq 1$. In this case $g_{N}=(N-1)/(N+1)$. In Table \ref{table4}, we have considered the Lorenz functions $L_{N}(p)$ for $N\in \{1,\ldots,20\}$ and have calculated values of the associated Gini coefficient $g_{N}$ and the associated $k$-index $k_{N}$.
 \begin{table}[!tbh]
 \centering
\caption{The family of Lorenz functions of the form $L_N(p)=p^{N}$ for all $p\in [0,1]$ where $N\in \{1,\ldots,20\}$.}
\label{table4}
\begin{tabular}{|c|c|c|c|}
\hline
$N$   &        $L_{N}(p)=p^N$  &     $g_{N}$   &        $k_{N}$  \\
\hline
$(N=1)$    &      $p$      &   0           &       $\frac{1}{2}$ \\
\hline
$(N=2)$   &       $p^2$         &   $\frac{1}{3}$          &       $\frac{\sqrt{5}-1}{2}\sim 0.618$ \\
\hline
$(N=3)$    &      $p^3$      &  $ \frac{1}{2}$   &  0.682\\
\hline
$(N=4)$    &      $p^4$      & $\frac{3}{5}$  &  0.725  \\
\hline
$(N=5)$    &      $p^5$      &   0.667   &          0.755  \\
\hline
$(N=6)$    &      $p^6$      & 0.714    &          0.778  \\
\hline
$(N=7)$    &      $p^7$     &  0.75  &          0.797  \\
\hline
$(N=8)$    &      $p^8$      &  0.778  &        0.812 \\
\hline
$(N=9)$    &      $p^9$     & 0.8       &       0.824\\
\hline
$(N=10)$    &      $p^{10}$      & 0.818  &     0.835  \\
\hline
\end{tabular}
\begin{tabular}{|c|c|c|c|}
\hline
$N$   &        $L_{N}(p)=p^N$  &     $g_{N}$   &        $k_{N}$  \\
\hline
$(N=11)$    &      $p^{11}$     &  0.833  &  0.844  \\
\hline
$(N=12)$    &      $p^{12}$      & 0.846  &  0.853 \\
\hline
$(N=13)$    &      $p^{13}$      &  0.857  &  0.860 \\
\hline
$(N=14)$    &      $p^{14}$      & 0.867  &  0.866  \\
\hline
$(N=15)$    &      $p^{15}$      & 0.875  &  0.872 \\
\hline
$(N=16)$    &      $p^{16}$      & 0.882  &  0.877 \\
\hline
$(N=17)$    &      $p^{17}$      & 0.889  &  0.882 \\
\hline
$(N=18)$    &      $p^{18}$      &  0.895  &  0.886 \\
\hline
$(N=19)$    &      $p^{19}$      &  0.9  &  0.890  \\
\hline
$(N=20)$    &      $p^{20}$      & 0.905  &  0.894  \\
\hline
\end{tabular}
\end{table}

In Table \ref{table4}, observe that if $N=13$, then $g_{13}\sim 0.857<k_{13}\sim 0.860$ and if $N=14$, then $g_{14}\sim 0.867>k_{14}\sim 0.866$ implying that coincidence takes place for some $N\in (13,14)$. Indeed, as demonstrated in Appendix ({\bf E}), the coincidence between the Gini coefficient and $k$-index occurs at some $N^*\in (N_1=13.82986,N_2=13.82987)$ and hence we have   
\begin{equation}
g_{N^*}=k_{N^*}\in \left(g_{N_1}=0.8651369,g_{N_2}=0.8651370\right)\in \left(\frac{6}{7},\frac{13}{15}\right).\hspace{1in}
\end{equation}Thus, we have established two things. First, there is no integer $N$ such that the Gini coefficient coincides with the $k$-index.  Second, if $N$ is any positive real number, then there exists an $N^*$ such that the Gini coincides with the $k$-index. 

\subsubsection{The second family} Suppose $L_{\alpha}(p)=1-(1-p)^{\alpha}$ for all $p\in [0,1]$ and given $\alpha\in (0,1)$. For this family of Lorenz functions, $g_{\alpha}=(1-\alpha)/(1+\alpha)$. Therefore, the value of $\alpha$ for which the Gini coefficient coincides with the $k$-index is obtained from $g_{\alpha}+L_{\alpha}(g_{\alpha})=1$ and computation shows that the solution is reached at $\alpha^*\simeq 0.072$ which implies that $g_{\alpha^*}=k_{\alpha^*}\simeq 0.865\in \left(6/7,13/15\right)$.  

\subsubsection{The hybrid family} Consider the hybrid family of Lorenz functions generated as the product of the two families of Lorenz function of this sub-section which is of the form $L_{(N\mid \alpha)}(p)=L_{N}(p)L_{\alpha}(p)=p^N(1-(1-p)^{\alpha})$ for all $p\in [0,1]$ with $N>1$ and $\alpha\in (0,1)$. In general, $g_{(N\mid \alpha)}=(N-1)/(N+1)-2B(N+1,\alpha+1)$ (Here $B(a,b)$ is the Beta function defined for $a>0$ and $b>0$). Suppose $\alpha=1/N$. Then for $N=1/\alpha=3.5$ we get $g_{(3.5\mid 1/3.5)}\simeq 0.806<k_{(3.5\mid 1/3.5)}\simeq 0.814$ and for $N=1/\alpha=4$ we get $g_{(4\mid 1/4)}\simeq 0.835>k_{(4\mid 1/4)}\simeq 0.830$. Therefore, at some $\overline{N}\in (3.5,4)$ we have $g_{(\overline{N}\mid 1/\overline{N})}=k_{(\overline{N}\mid 1/\overline{N})}\in (0.814,0.830)$ implying that $g_{(\overline{N}\mid 1/\overline{N})}=k_{(\overline{N}\mid 1/\overline{N})}\in \left(4/5,5/6\right)$.

\subsection{Finite polynomial Lorenz functions and the coincidence}
Consider any polynomial Lorenz function $L_{(n,a)}(p)=\sum_{m=1}^na_mp^m$ where $n$ is any positive integer greater than one and $a=(a_1,\ldots,a_n)$ is an $n$-element vector such that $a_m\geq 0$ for all $m=1,\ldots,n$ with at least one $a_m>0$. Then $L_{(n,a)}(1)=1$ implies that $\sum_{m=1}^na_m=1$. Therefore, each number $a_m$ is a weight with $a_m\in [0,1]$ for all $m\in \{1,\ldots,n\}$ and $\sum_{m=1}^na_m=1$. One can also verify that $L'_{(n,a)}(p)>0$ and $L''_{(n,a)}(p)>0$ for all $p\in (0,1)$. In Appendix {\bf F} we establish the following results:
\begin{enumerate}
	\item[{\bf F(a)}] \emph{If $n=2$, then, for any Lorenz function $L_{(2,a)}(p)=a_1p+(1-a_1)p^2$ with $a_1\in [0,1]$, we cannot have a coincidence between the Gini co-efficient and the $k$-index.}
	\item[{\bf F(b)}] \emph{If $n>2$ and if we have a coincidence between the Gini co-efficient and the $k$-index for some polynomila Lorenz function $L_{(n,a)}(p)$, then it is necessary that $g_{(n,a)}=k_{(n,a)}\geq \left(1/2\right)^{(1/3)}\sim 0.794.$} 
\end{enumerate}
{\bf F(a)} provides an impossibility result and, assuming coincidence between the Gini coefficient and the $k$-index and assuming $n>2$, {\bf F(b)} provides a necessary lower bound on the coincidence value. Even by restricting our analysis to finite polynomial Lorenz functions of the form $L_{(n,a)}(p)$, it is difficult to give an exact analytical argument on the upper bound on the coincidence between the Gini coefficient and the $k$-index. However, in what follows, we provide some selections of $\{\{a_m\}_{m=1}^n\}$ such that we can throw some light on the upper bound on the coincidence value. 

 \begin{table}[!tbh]
 \centering
\caption{Different polynomial Lorenz functions.}
\label{table5}
\begin{tabular}{|c|c|c|c|}
\hline
$Cases$   &        $L_{(n,a)}(p)$  &     $g_{(n,a)}=k_{(n,a)}$   &        ${\rm Interval} \ {\rm of} \ n$  \\
\hline
$(I)$    &      $p^n $      &   0.865           &       (13,14) \\
\hline
$(II)$   &       $\sum\limits_{m=1}^n\left(\frac{1}{n}\right)p^m$         &   0.869          &       (65,66) \\
\hline
$(III)$    &      $\left\{\frac{(n-1)}{n}\right\}p+\sum\limits_{m=2}^{n}\left\{\frac{1}{n(n-1)}\right\}p^m $      &  $ \rm Impossibility$   &  $--$\\
\hline
$(IV)$    &      $\sum\limits_{m=1}^{n-1}\left\{\frac{1}{n(n-1)}\right\}p^m + \left\{\frac{(n-1)}{n}\right\}p^{n} $      &   0.874  &  (17,18)  \\
\hline
$(V)$    &      $\sum\limits_{m=1}^{n}\left\{\frac{6m(n+1-m)}{n(n+1)(n+2)}\right\}p^m$      &   0.874   &         (40,41)  \\
\hline
$(VI)$    &      $\sum\limits_{m=1}^{n}\left\{\frac{2(m+1)}{n(n+3)}\right\}p^m$       & 0.877    &        (29,30)  \\
\hline
$(VII)$    &      $\sum\limits_{m=1}^{n}\left\{\frac{\ln(n+1-m)}{\sum\limits_{r=1}^n\ln(n+1-r)}\right\}p^m $     &  0.881  &          (77,78)  \\
\hline
\end{tabular}
\end{table}

In Table \ref{table5}, we have considered seven possibilities of finite polynomial Lorenz functions. The first possibility (Case (I)) is one where $a_1=\ldots=a_{n-1}=0$ and $a_n=1$ and this case is dealt with in Sub-section 1.2. Case (II) is one where $a_1=\ldots=a_n=1/n$ and we get a coincidence value of 0.869 (approximately) for some number lying between integers 65 and 66 and the coincidence value is higher than that under Case (I). Case (III) is one where $a_1=(n-1)/n$ and $a_2=\ldots=a_n=1/[n(n-1)]$ and since the weight $a_1$ is too high, $k$ is sufficiently close to $1/2$ and hence we do not get a coincidence between the Gini coefficient and the $k$ index. Case (IV) is one where $a_n=(n-1)/n$ and $a_1=\ldots=a_{n-1}=1/[n(n-1)]$ and we get the coincidence value of 0.874 (approximately) for some number lying between the integers 17 and 18 and this coincidence value is more than what we have for Case (II). In Case (V) we consider $a_m=[6m(n+1-m)]/[n(n+1)(n+2)]$ for all $m=\{1,2,\ldots,n\}$ and we get the coincidence value as 0.874 with some number lying in the interval 40 and 41 which is not an improvement in terms of coincidence value in comparison to Case (IV). In Case (VI) we consider $a_m=[2(m+1)]/[n(n+3)]$ for all $m=\{1,2,\ldots,n\}$ and we get the coincidence value as 0.877 for some number lying between the integers 29 and 30. The maximum coincidence value of $0.881$ is achieved for some number lying between integers 77 and 78 in Case (VII) where $a_m=\sum_{m=1}^n\left[\ln (n+1-m)/\left\{\sum_{r=1}^n\ln (n+1-r)\right\}\right]$ for all $m=\{1,2,\ldots,n\}$. From all these seven cases discussed in Table \ref{table5} it follows that the coincidence value is less than $8/9$. Observe that in all the cases dealt with in Table \ref{table5}, we cannot find a positive integer $n$ such that the Gini coefficient coincides with the $k$-index. Given the necessary condition {\bf F(b)} and given the results from Table \ref{table5}, our conjecture is that if the Gini coefficient coincides with the $k$-index for some nice polynomial Lorenz function of degree more than two, then the coincidence value must be more than $\left(1/2\right)^{(1/3)}\sim 0.7937$ and less than $8/9\sim 0.88\ldots$, that is, we must have $g_{(n,a)}=k_{(n,a)}\in \left(\left(1/2\right)^{(1/3)},8/9\right)$. Observe that in this subsection we could not obtain any finite polynomial Lorenz function for which we have a coincidence between the Gini coefficient and the $k$-index.

\subsection{Lorenz functions as limits of a tractable polynomial family} For any real number $b>1$, consider the associated finite polynomial Lorenz function given by
\begin{equation} \label{asymptote}
L_{(n,b)}(p)=\frac{\sum\limits_{r=1}^nb^{n-r}p^r}{\sum\limits_{r=1}^nb^{n-r}}.
\end{equation}In Appendix {\bf G} it is established that if the Lorenz function is given by (\ref{asymptote}), then, as $n\rightarrow \infty$, we have a non-trivial and symmetric Lorenz function $L_{b}(p)$ of the following form:
\begin{equation} \label{asymptote1}
L_{b}(p)=\frac{(b-1)p}{\left(b-p\right)}.
\end{equation}For the Lorenz function given by (\ref{asymptote1}), it is shown in Appendix ({\bf H}) that coincidence takes place for $b^*\simeq 1.022$ and we get the following coincidence value:  
\begin{equation}
g_{b^*}=k_{b^*}\simeq 0.873\in \left(\frac{13}{15},\frac{7}{8}\right). \hspace{2in}
\end{equation}

\subsection{Convergence towards coincidence with some simple Lorenz functions} In this sub-section we provide Table VI. to show how we reach towards the convergence conclusion between $g$ and $k$ by taking some simple Lorenz functions (also see Fig.~\ref{fig:lor_3}).

\begin{table}[!tbh]
\centering
\caption{Some simple Lorenz functions.}
\label{table6}
\resizebox{\columnwidth}{!}{%
\begin{tabular}{|c|c|c|c|}

\hline

$Cases$   &        $L(p)$  &     $g$   &        $k$  \\

\hline

$(i)$    &      $p$      &   0          &       $\frac{1}{2}$ \\

\hline

$(ii)$   &       $p^2$         &   $\frac{1}{3}$          &       $\frac{\sqrt{5}-1}{2}\sim 0.618$ \\

\hline

$(iii)$    &      $1-\sqrt{1-p}$      &  $\frac{1}{3}$   &  $\frac{\sqrt{5}-1}{2}\sim 0.618$\\

\hline

$(iv)$    &      $p+(1-p)\ln (1-p)$      &   $\frac{1}{2}$  &  0.682   \\

\hline

$(v)$    &      $1-\sqrt{1-p^2}$      &   $\frac{\pi}{2}-1\sim 0.571$  &         $\frac{1}{\sqrt{2}}\sim 0.707$ \\

\hline

$(vi)$    &      $=0, \ if \ p\in \left[0,\frac{1}{2}\right], \ =\left(p-\frac{1}{2}\right) \ if \ p\in \left[\frac{1}{2},1\right), \ and, \ =1 \ if \ p=1 $       & $\frac{3}{4}$    &        $\frac{3}{4}$  \\

\hline

$(vii)$    &      $=\left(\frac{1}{6}\right)\left(\frac{6p}{5}\right)^{4}, \ if \ p\in \left[0,\frac{5}{6}\right], \ and, \ =1-\frac{5}{6}\{6(1-p)\}^{\frac{1}{4}} \ if  \ p\in \left[\frac{5}{6},1\right] $     &  $\frac{5}{6}=0.833\ldots$  &          $\frac{5}{6}=0.833\ldots$  \\

\hline

$(viii)$    &      $=\left(\frac{1}{9}\right)\left(\frac{9p}{8}\right)^{\frac{25}{7}}, \ if \ p\in \left[0,\frac{8}{9}\right], \ and, \ =1-\frac{8}{9}\{9\left(1-p\right)\}^{\frac{7}{25}} \ if  \ p\in \left[\frac{8}{9},1\right]$     &  $\frac{8}{9}\sim 0.88\ldots $  &          $\frac{8}{9}\sim 0.88\ldots$  \\

\hline

$(ix)$    &      $0, \ if \ p\in [0,1), \  and \ 1 \ if \ p=1$     &  $1$  &    $k\rightarrow 1_{-}$  \\

\hline

\end{tabular}
}
\end{table}


Case (i) is $L(p)=p$ for all $p\in [0,1]$ which is the case of perfect egalitarian distribution of income. Case (ii) is $L(p)=p^2$ for all $p\in [0,1]$ where the Lorenz function is of a very simple quadratic form. Case (iii) is $L(p)=1-\sqrt{1-p}$ for all $p\in [0,1]$ and it is interesting to note that $g$ has the same value under both Case (ii) and Case (iii) and  $k$ also has the same value under both Case (ii) and Case (iii). In Case (iv), we have  $L(p)=p+(1-p)\ln (1-p)$ for all $p\in [0,1]$ that represents the Lorenz function associated with the exponential distribution. Case (v) is  $L(p)=1-\sqrt{1-p^2}$ for all $p\in [0,1]$ which represents the arc of a circle $(L(p)-1)^2+p^2=1$ as the Lorenz function. Case (vi) represents a piecewise linear Lorenz function for which we get the coincidence with $k=g=3/4$. The Lorenz function in Case (vi) is the same as the one given by (\ref{coincidence34}) and the coincidence result is established in Appendix {\bf B}. Case (vii) gives a special Lorenz function for which $k=g=5/6\sim 0.833$ and Case (viii) gives a special Lorenz function for which $k=g=8/9\sim 0.889$. Finally, Case (ix) represents a Lorenz function associated with the most unfair society where one agent possess all the income and in this case we have $g=1$ and $k$ approaching unity. We derive each of the results in Table \ref{table6} (that is, the derivations of $g$ and $k$ for Cases-(i)-(ix)) in Appendix {\bf I}.   

We have first argued that with symmetric Lorenz
functions, the range of coincidence values of $g$
and  $k$ indices is between $[3/4,1)$
and that we could identify a Lorenz function
(\ref{coincidence34}) with coincidence value
of 3/4  and a family of Lorenz functions
(\ref{coincidencepoint8onwards}) that takes all
coincidence values from the interval $[4/5,1)$.
The Lorenz functions given in (\ref{coincidence34})
and (\ref{coincidencepoint8onwards}) are not analytic
functions. We believe that with analytic functions, this
coincidence value is bounded above by $8/9\simeq
0.89$. This is argued with a sequence of steps. We
first identify an exponential family of Lorenz functions
from which we get one coincidence result and this
coincidence occurs at $0.864664$. Then we consider
two important families of Lorenz functions and a hybrid
family of Lorenz functions resulting from these two
important families and establish that if coincidence
occurs then it does not exceed 0.866. Then we look
for polynomial Lorenz functions and do not find any
coincidence result and we infer that if we get
coincidence result then it must necessarily be greater
than $\left(1/2\right)^{(1/3)}$ and with the help of a table
of different polynomial Lorenz functions we conjecture
that coincidence value cannot exceed $8/9$. Then we
provide a family of Lorenz functions that is obtained
from as a limit distribution of a polynomial family of
Lorenz functions and find the coincidence value to be
$0.873130$. Finally, we take some well-known and
simple Lorenz function and show how we start with
$g$ and $k$ indices that are sufficiently
different and slowly move towards the coincidence
between the two.

\section{Summary and discussions}
Social inequalities are ubiquitous. 
It has, therefore, been argued to be an emergent property of multi-component interacting socio-economic systems \cite{zhukov}.
Formal descriptions of such systems (wealth distributions, financial markets, cryptocurrencies, citations dynamics etc.) could be achieved through
the frameworks of self-organized criticality \cite{soc25,cite_soc} and also through nonextensive statistics \cite{tsallis2,lorenz_tsallis}. 
Nevertheless, quantification of such inequalities and their universal nature (first conjectured by Pareto's 80-20 law \cite{Pareto1971Translation}: implying $k=0.80$ in this context) remains an important 
and outstanding question. Here we attempt to quantify socio-economic inequality in a variety of systems that are predominantly highly competitive and are without 
external interventions designed to restrict emerging inequality among the participating agents. A complex dynamical system evolving without external fine tuning 
is precisely among the requirements of a self-organized critical system. Indeed, many of the systems studied here have been investigated through the construct of SOC 
systems (see e.g., for financial markets \cite{markets_soc}, citation evolution \cite{cite_soc}, crypocurrencies \cite{bitcoin_soc}, political behavior \cite{politics_soc} and so on). It is remarkable, therefore, that we observe the behavior of inequality indices (specifically, Gini $g$ and Kolkata $k$ indices) in this wide variety of socio-economic systems show nearly universal characteristics of approaching each other to a value of about $0.87$, which is precisely what is seen for the SOC models of physical systems \cite{mbc,front}.   

Specifically,
we have
analyzed here several data sets for
income distributions, fluctuations in the pricing of bitcoins, citation distributions of
some individual prize winning scientists, vote
share of the contestants in some elections. We analyzed here the income data (both for income and income
tax), in particular the IRS (USA) data
\cite{IRS,Ludwig2021} for the 36 year period from 1983 to 2018 (in section \ref{subsecA}), data for income from movie productions in both  Hollywood (USA \cite{Hollywood2011}) and Bollywood (India \cite{Bollywood2011}) during the 9 year period  2011-2019 (in section \ref{subsecA}), Google Scholar citations data for papers written by individual scientists having individual Google Scholar pages with ‘verified email’ addresses and have won Fields Medal (mathematics), or Boltzmann Medal (statistical physics), or ASICTP Dirac Medal (physics), or John von Neumann  Award (social science) in different years (in section \ref{subsecB}), and for vote share data of the candidates competing in parliament elections in India for the last two election years 2014 and 2019 \cite{Lokesabha2014,Lokesabha2019} (see section \ref{subsecB}). All these sectoral inequality analysis gives the index values for Gini ($g$) and Kolkata ($k$), and are plotted in Figs. \ref{fig2} and \ref{fig3}, and given in Tables \ref{table2} and \ref{table3}. The compiled Figs. \ref{fig6},\ref{fig7} from these isolated sectors of our society not only suggests the same pattern of growth of inequalities under competition, they clearly point towards the emerging coincidence of $g$ and $k$ indices at a universal value of about 0.87. 
In Fig. \ref{irs_gk}, we also compare  the growth pattern of the IRS (USA) income inequalities ($k$ and $g$ values) with that for the growth of citation inequalities of the papers published by the established universities, individual Nobel laureate scientists, and published in established journals (data taken from other publications, refs. \cite{Ghosh2021,Chatterjee2017}). Again, all these results of data analysis clearly indicate a dynamical drift of the inequality index values of Kolkata ($k$) and Gini ($g$) towards an universal inequality measure  $k = g = 0.87 \pm 0.02$ under unrestricted competitions. Note that by dynamics, we mean the long-term evolution and saturation of the aforementioned systems to their present state.
Further to support our findings that $g$ and $k$ indices tend to stabilize around a value of 0.87, we show (in Fig~\ref{fig3}) the $g$ and $k$ index values for the daily price fluctuation distributions for Bitcoin, calculated for a decade (2010-2021). 
It clearly demonstrates that in absence of any central bank to control the price fluctuations of any crypto-currency (unlike the different national currencies) both the inequality indices ($g$ and $k$) approach a value around $0.87$ and then fall down subsequently. Generally, $g$ and $k$ values corresponding to daily Bitcoin price fluctuations do not cross this limiting value ($g = k \simeq 0.87$). We then discussed in section \ref{sec3} some general analytical and structural features of the Lorenz function $L(p)$, and for the bounds for equality of  $g$ and $k$ indices we argued (see Table \ref{table4}) that for most reasonable analytic forms of $L(p)$ the coincidence values of  Gini ($g$)  and  Kolkata ($k$) indices will lie between 4/5 (= 0.80)  and 8/9 (= 0.88\dots).

It is interesting to note that for the IRS data, the inequality indices have not quite reached the value 0.87 (that we argue here to be the attractor resulting from an SOC state), but Fig. \ref{irs_gk} in Sec.2.5 clearly shows that both $g$ and $k$ are consistently growing with time, presumably due to gradual withdrawal of public welfare programs in the US and  have already crossed the Pareto value $k=0.80$. It may be predicted to reach 0.87 with the complete withdrawal of the above mentioned public welfare programs i.e., allowing the participants to reach a state of unrestricted competition.

Generally speaking, our study (both data analysis and
mathematical structural analysis) here clearly confirms
that while the Gini index ($g$) and Kolkata index ($k$)
could in principle  assume any value from zero or half
respectively (for perfect equality) to  unity (for both
the indices for extreme inequality), the social
dynamics of competitions take the  index values of
$g = k \simeq 0.87$, suggesting that about 87\% wealth,
citations or votes are possessed, earned or won by 13\%
people, papers or election candidates in cases of unrestricted competition. This may be
a quantitative and generalized (over all social
sectors) version of the more than a century old 80-20
(or $k = 0.80$) law of Pareto \cite{Pareto1971Translation}. 
In fact, this feature of the inequality indices stems from the SOC nature of the underlying dynamics and have been shown to
 be present in a myriad of SOC models in physical science \cite{mbc,front}. 

\appendix
\section*{Appendices:}
\subsection{Appendix {\bf A}:}  
In Figure \ref{fig:symmetriclc}, the area between $45^0$ line and the Lorenz curve (shaded in red) is $g/2$. Define the area of the rectangle $AF_1F_2C=AC\times OP$ as $\mathcal{A}$. It is easy to work out that $\mathcal{A}=AC\times OP=\{(1/\sqrt{2})-\sqrt{2}(1-k)\}\times \sqrt{2}=(2k-1)$. Suppose, we take $g=k$ and $g/2=\alpha \mathcal{A}$ for some fraction $\alpha$. Then, on simplification we have 
\begin{equation}\label{coincidencealpha}
g=k=\frac{2\alpha}{(4\alpha-1)}.
\end{equation}

\begin{figure}[!tbh]
    \centering
    \includegraphics[scale=0.67]{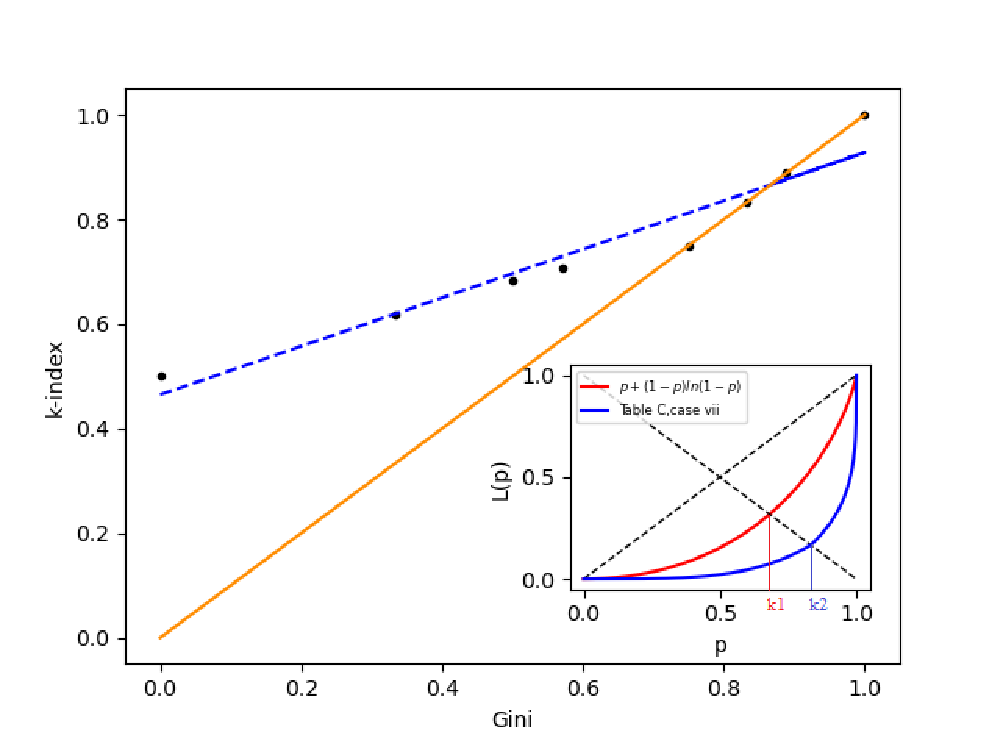}
    \caption{Gini ($g$) vs. $k$-index ($k$) graph. Here Orange line represents the $g = k$ line. Black dots represent ($g,k$) values for some simple Lorenz functions given by Table \ref{table6}. The trend of the black dots could be seen to converge towards $g=k$ line. Blue dashed line depicts the best-fit line. In inset two different Lorenz curves are shown for cases (iv) and (vii) from Table \ref{table6}. Red curve is the Lorenz curve for case (iv) and $k$-index value, k1 = 0.682 and similarly the blue curve represents the Lorenz curve for case (vii) and $k$-index value, k2 = 0.833.}
    \label{fig:lor_3}
\end{figure}

\begin{figure}[!tbh]
    \centering
    \includegraphics[scale=0.95]{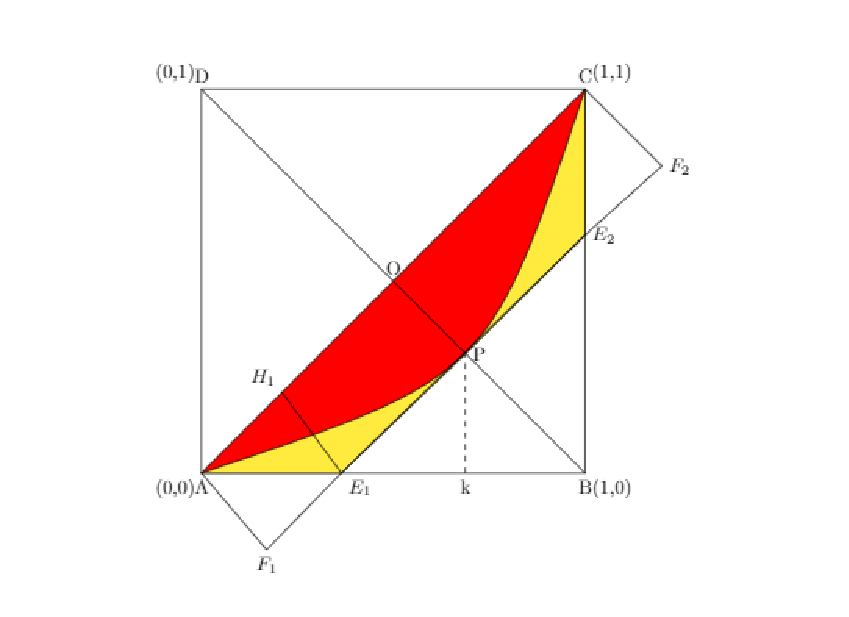}
    \caption{The Gini coefficient ($g$) coincides with the $k$-index ($k$) and the coincidence value must necessarily belong to the interval $[3/4,1)$, that is, $k=g\in [3/4,1).$ }
   \label{fig:symmetriclc}
\end{figure}

First note that from condition (\ref{coincidencealpha}), we have $\alpha=1/2$ if and only if $g=k=1$. However, due to non-triviality of the Lorenz function, $\alpha=1/2$ is ruled out. Therefore, given $k,g< 1$ and given that the function $r(\alpha):=2\alpha/(4\alpha-1)$ is decreasing in $\alpha$, for coincidence to take place we must have {\bf (i)} $\alpha>1/2$. Moreover, note that the maximum area that $g/2$ can potentially have is when it coincides with the trapezium $AE_1E_2C$ (which is the sum of areas shaded in red and yellow). Using symmetry of the Lorenz function, the area of this trapezium $AE_1E_2C={\mathcal{A}}-\triangle AF_1E_1-\triangle  CF_2E_2={\mathcal{A}}-\square AF_1E_1H_1={\mathcal{A}}-(H_1E_1)^2={\mathcal{A}}-(OP)^2=\mathcal{A}-(2k-1)^2/2=\mathcal{A}(3-2k)/2$. Hence, the maximum value that $\alpha$ can take is $\alpha_{M}=(3-2k)/2$. Therefore, given $\mathcal{A}=2k-1$ and $\alpha_M=(3-2k)/2$, if $g$ coincides with $k$, then we must have $k/2\leq \alpha_M\mathcal{A}$ which means that if $g$ coincides with $k$, then we must necessarily have the following inequality:
 \begin{equation}\label{thikapproach}
 \frac{k}{2}\leq \left(\frac{3-2k}{2}\right)\times (2k-1).
 \end{equation} From (\ref{thikapproach}) it follows that 
 \begin{equation}
 4\times \left(k-\frac{3}{4}\right)\times \left(k-1\right)\leq 0\Rightarrow \ g=k\in \left[\frac{3}{4},1\right]. 
 \end{equation}Further, given $g=k=1$ is ruled out (due to {\bf (i)}), we must have $k=g\in \left[\frac{3}{4},1\right)$. 
 
\subsection{Appendix {\bf B}} Consider the Lorenz function given by $L_{\left(\frac{3}{4}\right)}(p)$ in (\ref{coincidence34}). For any $p\in (1/2,1)$, $L_{\left(\frac{3}{4}\right)}(p)>0$ and hence the Lorenz function $L_{\left(\frac{3}{4}\right)}(p)$ is non-trivial. Moreover, for any $p\in (1/2,1)$, $L_{\left(\frac{3}{4}\right)}(1-L_{\left(\frac{3}{4}\right)}(p))=L_{\left(\frac{3}{4}\right)}(1-\{p-(1/2)\})=L_{\left(\frac{3}{4}\right)}((3/2)-p)=\{(3/2)-p\}-(1/2)=1-p$ implying symmetry. Note that $L_{\left(\frac{3}{4}\right)}(p)=0$ for all $p\in [0,1/2]$, $L_{\left(\frac{3}{4}\right)}(p)$ is continuous at $p=1/2$ and $L_{\left(\frac{3}{4}\right)}(p)$ is increasing with a constant slope of unity for all $p\in (1/2,1)$ and hence the function $L_{\left(\frac{3}{4}\right)}(p)$ is non-decreasing, continuous and convex in $(0,1)$. Observe that $L_{\left(\frac{3}{4}\right)}(3/4)=(3/4)-(1/2)=1/4$ and hence $L_{\left(\frac{3}{4}\right)}(3/4)+3/4=1$ which means that $k_{\left(\frac{3}{4}\right)}=3/4$. Further, $g_{\left(\frac{3}{4}\right)}=1-2\int_{0}^1L_{\left(\frac{3}{4}\right)}(P)dP=1-2\int_{1/2}^1L_{\left(\frac{3}{4}\right)}(P)dP=1-2\{(1/2)p^2\}_{p=1/2}^{p=1}+(1/2)=(3/2)-2\{(3/8)\}=3/2-3/4=3/4$. Hence, we have $g_{\left(\frac{3}{4}\right)}=k_{\left(\frac{3}{4}\right)}=3/4=0.75$.
 
\subsection{Appendix {\bf C}} For any rational fraction $q\in [4/5,1)$, consider the associated Lorenz function given by $L_{(q)}(p)$ in (\ref{coincidencepoint8onwards}). For any $p\in (1-q,1)$, $L_{(q)}(p)>0$ and hence the Lorenz function $L_{(q)}(p)$ is non-trivial. Define $y_1(p):=(1-q)\left(\frac{p+q-1}{2q-1}\right)^{3}$. Observe that for any $p\in [1-q,q]$,
 \begin{equation}\label{ek}
 L_{(q)}\left(1-L_{(q)}(p)\right)=q-(2q-1)\left(\frac{1-\{1-y_1(p)\}}{1-q}\right)^{\frac{1}{3}}=q-(p+q-1)=1-p.
 \end{equation}Similarly, define $y_2(p):=q-(2q-1)\left(\frac{1-p}{1-q}\right)^{\frac{1}{3}}$ and note that for any $p\in [q,1)$,
 \begin{equation}\label{dui}
L_{(q)}\left(1-L_{(q)}(p)\right)=(1-q)\left(\frac{\{1-y_2(p)\}+q-1}{2q-1}\right)^{3}=(1-q)\left(\frac{1-p}{1-q}\right)=1-p.
 \end{equation}From (\ref{ek}) and (\ref{dui}), it follows that for any rational fraction $q\in [4/5,1)$, the associated Lorenz function given by (\ref{coincidencepoint8onwards}) is symmetric. Further, observe that  (i) $L_{(q)}(p)=0$ for all $p\in [0,1-q]$, (ii) $\lim_{p\rightarrow q_-}L_{(q)}(p)=\lim_{p\rightarrow q_+}L_{(q)}(p)=(1-q)$, (iii) $L'_{(q)}(p)=3(1-q)(p+q-1)^2/\left\{(2q-1)^3\right\}>0$ for all $p\in (1-q,q)$ and (iv) $L'_{(q)}(p)=(2q-1)(1-p)^{\frac{1}{3}-1}/\left\{3(1-q)^{\frac{1}{3}}\right\}>0$ for all $p\in (q,1)$. Condition (i)-(iv) together imply that for any rational fraction $q\in [4/5,1)$, the associated Lorenz function $L_{(q)}(p)$ given by (\ref{coincidencepoint8onwards}) is non-decreasing and continuous in $(0,1)$. Moreover, (I) $\lim_{p\rightarrow q_-}L'_{(q)}(q)\leq \lim_{p\rightarrow q_+}L_{(q)}(q)$ for all $q\in [4/5,1)$, (II) $L''_{(q)}(p)=6(1-q)(p+q-1)/\left\{(2q-1)^3\right\}>0$ for all $p\in (1-q,q)$ and (III) $L''_{(q)}(p)=2(2q-1)(1-p)^{\frac{1}{3}-2}/\left\{9(1-q)^{\frac{1}{3}}\right\}>0$ for all $p\in (q,1)$. From (I)-(III) it follows that for any rational fraction $q\in [4/5,1)$, the associated Lorenz function $L_{(q)}(p)$ given by (\ref{coincidencepoint8onwards}) is convex in $(0,1)$. To show coincidence, firstly note that  $L_{(q)}(q)=1-q$ implies that $k_{(q)}=q$. Finally, observe that   $g_{(q)}=1-2\int_{0}^1L_{(q)}(P)dP=1-2\left\{\frac{(1-q)(2q-1)}{2}+(1-q)^2\right\}=1-2\left\{\frac{(1-q)}{2}\right\}=q=k_{(q)}$. Thus, for any rational fraction $q\in [4/5,1)$ with the associated Lorenz function $L_{(q)}(p)$ given by (\ref{coincidencepoint8onwards}), we have $g_{(q)}=k_{(q)}=q$. 
 
\subsection{Appendix {\bf D}} If $L_{\beta}(p)=\frac{(e^{\beta p}-1)}{(e^{\beta}-1)}$ for all $p\in [0,1]$ with $\beta>0$. Then $$g_{\beta}=1-\frac{2[e^{\beta}-(1+\beta)]}{\beta(e^{\beta}-1)}=\frac{(2+\beta)-(2-\beta)e^{\beta}}{\beta(e^{\beta}-1)}=\left(\frac{\beta-2}{\beta}\right)+\left(\frac{2}{e^\beta-1}\right).$$Hence, we have $g_{\beta}=\left\{\left(\beta-2\right)/\beta\right\}+\left\{2/\left(e^\beta-1\right)\right\}$. Moreover, observe that for the $k$-index $k_{\beta}$ to coincide with the Gini coefficient $g_{\beta}$ we must find an $\beta>0$ such that  
\noindent
$\frac{e^{\beta k_\beta}-1}{e^{\beta}-1}=1-g_{\beta}=\frac{2[e^{\beta}-(1+\beta)]}{\beta(e^{\beta}-1)}$
\noindent
$\Leftrightarrow e^{\beta k_\beta}-1=\frac{2[e^{\beta}-(1+\beta)]}{\beta}$
\noindent
$\Leftrightarrow  e^{\beta k_\beta}=1+\frac{2[e^{\beta}-(1+\beta)]}{\beta}$
\noindent
$\Leftrightarrow  e^{\beta k_\beta}=\frac{[2e^{\beta}-2-\beta]}{\beta}$
\noindent
$\Leftrightarrow  k_{\beta}=\frac{\ln [2e^{\beta}-2-\beta]-\ln(\beta)}{\beta}$
\noindent
$\Leftrightarrow \frac{(2+\beta)-(2-\beta)e^{\beta}}{\beta(e^{\beta}-1)}=\frac{\ln [2e^{\beta}-2-\beta]-\ln(\beta)}{\beta}$
\noindent
$\Leftrightarrow (2+\beta)-(2-\beta)e^{\beta}=(e^{\beta}-1)\left\{\ln (2e^{\beta}-2-\beta)-\ln(\beta)\right\}$
\noindent
$\Leftrightarrow \beta(e^{\beta}+1)-2(e^{\beta}-1)=(e^{\beta}-1)\left\{\ln ([2e^{\beta}-2-\beta)-\ln(\beta)\right\}$
\noindent
$\Leftrightarrow \beta(e^{\beta}+1)=(e^{\beta}-1)\left\{2+\ln (2e^{\beta}-2-\beta)-\ln(\beta)\right\}$
\noindent
$\Leftrightarrow 2\beta=(e^{\beta}-1)\left\{(2-\beta)+\ln (2e^{\beta}-2-\beta)-\ln(\beta)\right\}.$ Hence, we get the following equation:
\begin{equation}\label{one}
2\beta-(e^{\beta}-1)\left\{(2-\beta)+\ln (2e^{\beta}-2-\beta)-\ln(\beta)\right\}=0.
\end{equation}The only solution to (\ref{one}) is at $\beta^*=14.778$. Hence, we have $g_{\beta^*}=k_{\beta^*}=0.865$.

\subsection{Appendix {\bf E}:}
Coincidence between the Gini coefficient and the $k$-index can take place for some $N>1$ if and only if  
\noindent
$\left(\frac{N-1}{N+1}\right)+\left(\frac{N-1}{N+1}\right)^{N}=1$
\noindent
$\Leftrightarrow (N-1)^{N}=2(N+1)^{(N-1)}$
\noindent
$\Leftrightarrow \left(\frac{N+1}{N-1}\right)^{N-1}=\frac{(N-1)}{2}$
\noindent
$\Leftrightarrow \left(1+\frac{2}{N-1}\right)^{N-1}=\frac{(N-1)}{2}$
\noindent
$\Leftrightarrow \left(1+\frac{1}{\frac{N-1}{2}}\right)^{N-1}=\frac{(N-1)}{2}$
\noindent
$\Leftrightarrow \left(1+\frac{1}{\frac{N-1}{2}}\right)^{\frac{N-1}{2}}=\left(\frac{(N-1)}{2}\right)^{\frac{1}{2}}$. 
Therefore, for the coincidence it is necessary and sufficient that 
\begin{equation} \label{a}
H(y(N)):=\left\{\left(1+\frac{1}{y(N)}\right)^{y(N)}-\left(y(N)\right)^{\frac{1}{2}}\right\}=0, \ {\rm where} \ y(N):=\frac{N-1}{2}.
\end{equation}
Observe that $H(y(2))=H(1/2)=\sqrt{3}-1/\sqrt{2}=1.025>0$ and if $H(y(2e^2+1))=H(e^2)$, then $H(e^2)=-0.164<0$. Moreover, observe that $H(y(N))<0$ for all $y(N)\in (e^2,\infty)$ since for any $y(N)\in (e^2,\infty)$, we have $(1+(1/y(N)))^{y(N)}\leq e$ and $\sqrt{y(N)}>e$. By plotting the function $H(y(N))$ for all $y(N)$'s lying in the inetrval $(1/2,e^2]$, one can verify that there exists a unique $y(N^*)\in (1/2,e^2]$ such that $H(y(N^*))=0$. Define $N_1:=13.82986$ and $N_2:=13.82987$ and observe that $H(y(N_1))=H(6.41493)=0.00000127>0$, and $H(y(N_2))=H(6.41494)=-0.00000041<0$. Therefore, it is obvious that $y(N^*)\in (6.41493,6.41494)$, that is $N^*\in (N_1,N_2)$. Moreover, $g_{N_1}=0.8651369$ and $g_{N_2}=0.8651370$ implying that the Gini coefficient coincides with the $k$-index for $N^*\in (N_1,N_2)$ and $g_{N^*}=k_{N^*}\in (g_{N_1},g_{N_2})$.
 
\subsection{Appendix {\bf F}}If $n=2$, it is necessary that $a_2=(1-a_1)$ so that $L_{(2,a)}(p)=a_1p+(1-a_1)p^2$ and $g_{(2,a)}=(1-a_1)/3$. For Gini coefficient to coincide with $k$-index, it is also necessary that $3(1-a_1)+3a_1(1-a_1)+(1-a_1)^3=9$. However, this equation has no solution such that $a_1\in [0,1]$. Hence, if the Lorenz function is a polynomial of order two, then we cannot have a situation where the Gini coefficient coincides with the $k$-index. This argument established {\bf F(a)}.  Therefore, for the coincidence of Gini and $k$-index, we must have $n>2$ and for the coincidence, we must also satisfy the following requirement: 
\vspace{.1in}
\noindent
$g_{(n,a)}=1-\sum\limits_{m=1}^na_m\left(\frac{2}{m+1}\right)=1-L_{(n,a)}\left(g_{(n,a)}\right)=1-\sum\limits_{m=1}^na_m(g_{(n,a)})^m$
\noindent
$\Leftrightarrow \sum\limits_{m=1}^na_m\left\{\left(g_{(n,a)}\right)^m-\left(\frac{2}{m+1}\right)\right\}=0$.
From the last equation it follows that there exists at least one $m\in \{1,2,\ldots, n\}$ such that
\begin{equation}\label{necessary} \left(g_{(n,a)}\right)^m>\left(\frac{2}{m+1}\right) \ \ {\rm implying} \ \  g_{(n,a)}>\min\left\{\left(\frac{2}{m+1}\right)^{\frac{1}{m}}\right\}_{m=1}^n=\left(\frac{1}{2}\right)^{\frac{1}{3}}\sim 0.794.
\end{equation}
Hence, we have {\bf F(b)}.

\subsection{Appendix {\bf G}} For any real number $b>1$, suppose consider the Lorenz function given by $L_{(n,b)}(p)=\left\{\sum_{r=1}^nb^{n-r}p^r\right\}/\left\{\sum_{r=1}^nb^{n-r}\right\}$ and define the Lorenz function $L_b(p):=(b-1)p/(b-p)$ for all $p\in [0,1]$. Then we have 
\begin{equation} \label{asymptote3}
L_{(n,b)}(p)=\frac{\sum\limits_{r=1}^nb^{n-r}p^r}{\sum\limits_{r=1}^nb^{n-r}}=\frac{b^n(b-1)}{(b^n-1)}\sum\limits_{r=1}^n\left(\frac{p}{b}\right)^r=\frac{(b-1)}{\left(1-\left(\frac{1}{b}\right)^n\right)}\left(\frac{p}{b}\right)\left(\frac{1-\left(\frac{p}{b}\right)^n}{1-\left(\frac{p}{b}\right)}\right).
\end{equation}From (\ref{asymptote3}) it follows that as $n\rightarrow \infty$, $(1/b)^n\rightarrow 0$ and $(p/b)^n\rightarrow 0$ and we have the Lorenz function $L_{(\infty,b)}$ with the following form:
\begin{equation} \label{asymptote4}
L_{(\infty,b)}(p)=\frac{(b-1)}{\left(1-\left(\frac{p}{b}\right)\right)}\left(\frac{p}{b}\right)=\frac{(b-1)p}{\left(b-p\right)}=L_b(p).
\end{equation}Observe that given (\ref{asymptote4}), for any $b>1$, we have $L_{b}(0)=0$, $L_{b}(1)=1$, the first derivative $L'_{b}(p)=(b-1)b/(b-p)^2>0$, and, the second derivative $L''_{b}(p)=2(b-1)/(b-p)^3>0$. Thus, $L_{b}(p)$ is a well-defined Lorenz function for any given real number $b>1$. Moreover, since $L_{b}(p)>0$ for all $p\in (0,1)$, the Lorenz function $L_{b}(p)$ given by (\ref{asymptote4}) is non-trivial. Further, for any $p\in (0,1)$, $L_{b}(1-L_{b}(p))=L_{b}\left(\frac{b(1-p)}{b-p}\right)=\left(\frac{(b-1)\left(\frac{b(1-p)}{b-p}\right)}{b-\left(\frac{b(1-p)}{b-p}\right)}\right)=\left(\frac{(b-1)b(1-p)}{b(b-p)-b(1-p)}\right)=\left(\frac{(b^2-b)(1-p)}{b^2-b}\right)=1-p$. Hence, for any $p\in (0,1)$, we have established that $L_{b}(1-L_{b}(p))=1-p$. Thus, the Lorenz function $L_{b}(p)$ given by (\ref{asymptote4}) satisfies symmetry.  

\subsection{Appendix {\bf H}} On the one hand, coincidence requires that $g_{b)}+L_{b}(g_{b})=1$ which gives 
\begin{equation}\label{solve1}
g_{b}=b-\sqrt{b(b-1)}.
\end{equation}On the other hand, using $\int_{0}^1L_{b}(p)dp=b(b-1)\ln(b/(b-1))-(b-1)$ and the formula $g_{b}=1-2\int_{0}^1L_{b}(p)dp$, we get 

\begin{equation}\label{solve2}
g_{b}=1-2b(b-1)\ln\left(\frac{b}{b-1}\right)+2(b-1).
\end{equation} Solving for $b$ from (\ref{solve1}) and (\ref{solve2}) we get $b^*\simeq 1.022$ and therefore it easily follows that $g_{b^*}=k_{b^*}\simeq 0.873$.

\subsection{Appendix {\bf I}} {\it Case (i):} If $L(p)=p$ for all $p\in [0,1]$, then $\int_0^1L(P)dP=1/2$ so that $g=1-2\int_0^1L(P)dP=1-1=0$ and $k+L(k)=2k=1\Rightarrow k=1/2$.  
\noindent
{\it Case (ii):} If $L(p)=p^2$ for all $p\in [0,1]$, then $\int_0^1L(P)dP=1/3$ so that $g=1-2\int_0^1L(P)dP=1-2/3=1/3$ and $k+L(k)=k+k^2=1\Rightarrow k=(\sqrt{5}-1)/2\sim 0.618$. 
\noindent
{\it Case (iii):} If $L(p)=1-\sqrt{1-p}$ for all $p\in [0,1]$, then $\int_0^1L(P)dP=1-(2/3)$ so that $g=1-2\int_0^1L(P)dP=1-2+(4/3)=1/3$ and $k+L(k)=k+1-\sqrt{1-k}=1\Rightarrow k+k^2=1\Rightarrow k=(\sqrt{5}-1)/2\sim 0.618$.
\noindent
{\it Case (iv):} If $L(p)=p+(1-p)\ln (1-p)$ for all $p\in [0,1]$, then $\int_0^1L(P)dP=(1/2)-(1/4)=1/4$ so that $g=1-2\int_0^1L(P)dP=1-(1/2)=1/2$ and one can numerically verify that $k+L(k)=1\Rightarrow e^{(1-2k)}=(1-k)^{1-k}\Rightarrow k\sim 0.682$.
\noindent
{\it Case (v):} If $L(p)=1-\sqrt{1-p^2}$ for all $p\in [0,1]$, then $\int_0^1L(P)dP=1-(\pi/4)$ so that $g=1-2\int_0^1L(P)dP=1-2+(\pi/2)=(\pi/2)-1\sim 0.571$ and $k+L(k)=k+1-\sqrt{1-k^2}=1\Rightarrow 2k^2=1\Rightarrow k=(1/\sqrt{2})\sim 0.707$.
\noindent
{\it Case (vi):} This result is already established in Appendix {\bf B}. 
\noindent
{\it Case (vii):} Firstly, at $p=5/6$, we have $L(5/6)=1/6$ and hence $(5/6)+L(5/6)=1$ implying $k=5/6$. Moreover, observe that $\int_{0}^{5/6}L(P)dP=\left(1/36\right)$ and  $\int_{5/6}^1L(P)dP=\left(2/36\right)$. Hence, $\int_{0}^1L(P)dP=1/12$ and $g=1-(2/12)=5/6=k$. 
\noindent
{\it Case (viii):} Firstly, at $p=8/9$, we have $L(8/9)=1/9$ and hence $(8/9)+L(8/9)=1\Rightarrow k=8/9$. Moreover, observe that $\int_{0}^{8/9}L(P)dP=\left(7/324\right)$ and  $\int_{8/9}^1L(P)dP=\left(11/324\right)$. Hence, $\int_{0}^1L(P)dP=1/18$ implying $g=1-(2/18)=8/9=k$. 
\noindent
{\it Case (ix):} If $L(p)=0$ for $p\in [0,1)$ and $L(1)=1$, then $\int_0^1L(P)dP=0$ so that $g=1-2\int_{0}^1L(P)dP=1-0=1$ and since for every $p\in [0,1)$, $L(p)=0$, we have $k\rightarrow 1_{-}$.

\section*{acknowledgement}
We are extremely grateful to Victor Yakovenko and Danial Ludwig for providing us with the IRS Income and Tax data sets (US) for the period 1983 to 2018. SB would like to thank the DST, Government of India, for financial support of INSPIRE fellowship. BKC is thankful to the Indian National Science Academy for their Senior Scientist Research Grant.




\end{document}